\pdfoutput=1

\documentclass[%
    aip                ,   
    apl                ,   
    reprint            ,   
    showpacs           ,   
    superscriptaddress ,   
    nofootinbib            
]{revtex4-1}
\usepackage[T1]{fontenc}
\usepackage[utf8]{inputenc}
\usepackage[english]{babel}
\usepackage{afterpage}

\usepackage{amsmath}
\usepackage{amssymb}
\usepackage{braket}

\usepackage[hyperref]{xcolor}
\usepackage{graphicx}
\usepackage[export]{adjustbox}
\graphicspath{{figures/PNG/}{figures/PDF/}{figures/}}

\usepackage[
    unicode           = true  ,
    plainpages        = false , 
    pdfpagelabels     = true  , 
    bookmarks         = true  ,
    bookmarksnumbered = true  ,
    bookmarksopen     = true  ,
    breaklinks        = true  ,
    backref           = false ,
    colorlinks        = true  ,
    linkcolor         = blue  ,
    urlcolor          = blue  ,
    citecolor         = red   ,
    anchorcolor       = green ,
    hyperindex        = true  ,
    linktocpage       = true  ,
    hyperfigures      = true
]{hyperref}
\hypersetup{
    linkcolor         = [RGB]{000 114 189} ,  
    urlcolor          = [RGB]{000 114 189} ,  
    citecolor         = [RGB]{217 083 025} ,  
    anchorcolor       = [RGB]{119 171 048} ,  
    pdftitle={Spatial and Spectral Mode-Selection Effects in Topological Lasers with Frequency-Dependent Gain},
    pdfauthor={Matteo Seclì <matteo.secli@sissa.it>}
}

\def\supplementfilename{extra/topolaserTLA_suppl}
\newcommand{\suppref}[1]{\hyperref[sec:supp]{#1}}
\newcommand{\suppmat}{\suppref{supplementary material}}

\usepackage{ifxetex}
\usepackage{pdfpages}
\ifxetex\else\usepackage{pax}\fi
\usepackage{pgffor}
\makeatletter\AtBeginDocument{\let\LS@rot\@undefined}\makeatother
\ifxetex\newcount\pdflastximagepages\def\pdfximage#1{\pdflastximagepages=\XeTeXpdfpagecount"#1"\relax}\fi
\pdfximage{\supplementfilename.pdf}
\def\numbersupplementpages{\the\pdflastximagepages}
\newcommand{\includesupplementary}{
    \onecolumngrid
    \newpage
    \pdfbookmark[0]{Supplementary Information}{suppl}
    \foreach \x in {1,...,\numbersupplementpages}
    {
        \includepdf[pages=\x]{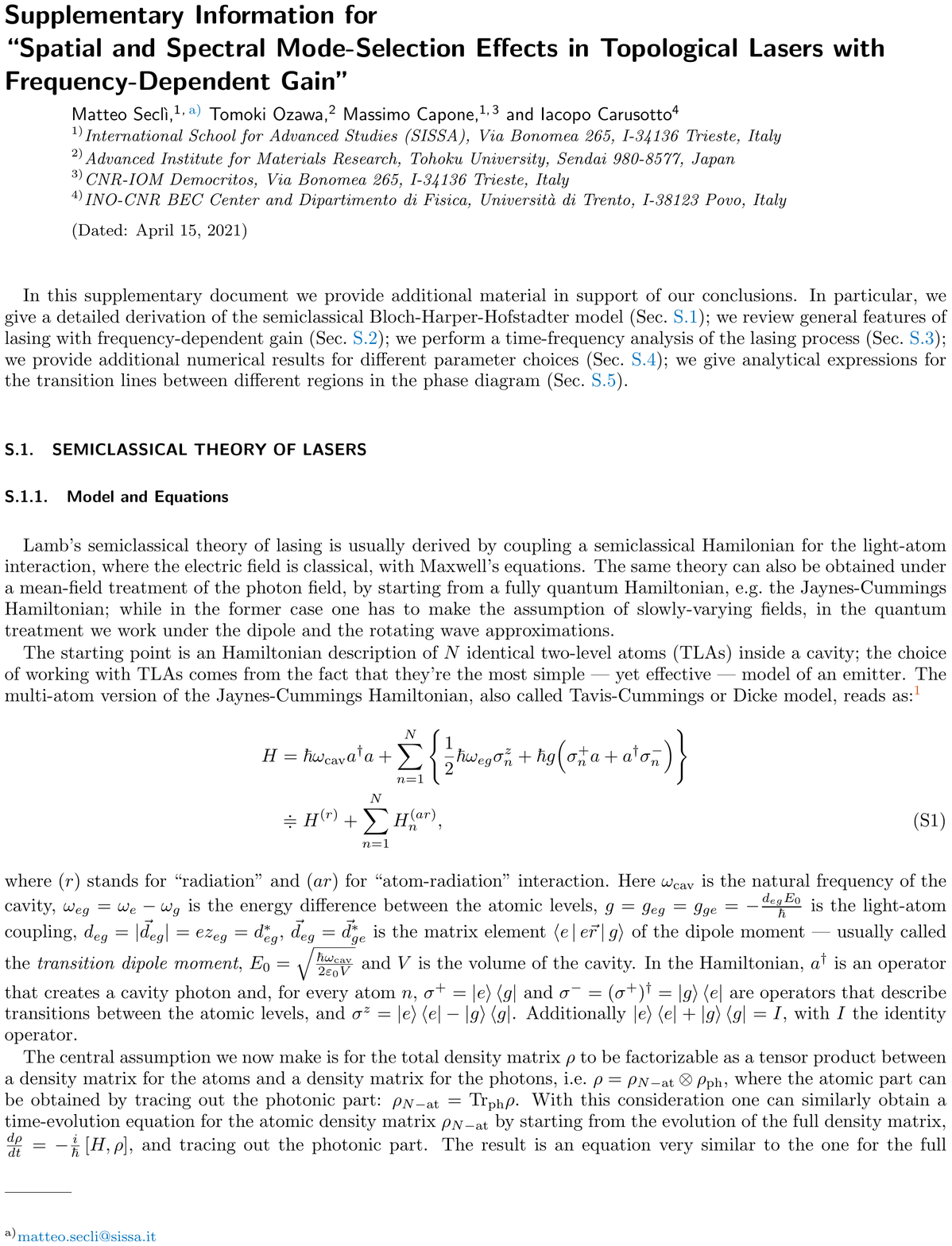}
    }
}

\begin{document}

\title{Spatial and Spectral Mode-Selection Effects in Topological Lasers with Frequency-Dependent Gain}

\author{Matteo Seclì}
\email[]{matteo.secli@sissa.it}
\affiliation{\mbox{International School for Advanced Studies (SISSA), Via Bonomea 265, I-34136 Trieste, Italy}}

\author{Tomoki Ozawa}
\affiliation{\mbox{Advanced Institute for Materials Research, Tohoku University, Sendai 980-8577, Japan}}

\author{Massimo Capone}
\affiliation{\mbox{International School for Advanced Studies (SISSA), Via Bonomea 265, I-34136 Trieste, Italy}}
\affiliation{CNR-IOM Democritos, Via Bonomea 265, I-34136 Trieste, Italy}

\author{Iacopo Carusotto}
\affiliation{\mbox{INO-CNR BEC Center and Dipartimento di Fisica, Università di Trento, I-38123 Povo, Italy}}

\date{April 15, 2021}

\begin{abstract}
We develop a semiclassical theory of laser oscillation into a chiral edge state of a topological photonic system endowed with a frequency-dependent gain. As an archetypal model of this physics, we consider a Harper-Hofstadter lattice embedding population-inverted two-level atoms as gain material. We show that a suitable design of the spatial distribution of gain and of its spectral shape provides flexible mode selection mechanisms that can stabilize single-mode lasing into an edge state. Implications of our results for recent experiments are outlined.
\end{abstract}


\maketitle

\section{Introduction}

Topological lasers (in short, \textit{topolasers}) are one of the most promising applications of topological photonics. Such devices are obtained by including a suitable gain material in a topological system so to induce laser oscillation in a topological edge state.~\cite{Ozawa2019,Ota2020} Stimulated by pioneering theoretical proposals,~\cite{Harari2016,Wittek2017,Harari2018,Pilozzi2016,Solnyshkov2016} experimental realizations were first reported for the zero-dimensional edge states of one-dimensional arrays.~\cite{St-Jean2017,Parto2018,Han2019,Ota2018,Zhao2018} Extension to nanolasers based on zero-dimensional corner states of two-dimensional lattices was reported in Ref.~\onlinecite{Zhang2020,Kim2020}.

Scaling up in dimension, the crucial advantages for optoelectronic applications offered by topological lasing into the one-dimensional edge modes of a two-dimensional lattice have been theoretically highlighted:~\cite{Harari2016,Wittek2017,Harari2018,Kartashov2019,Amelio2019} the topological protection of chirally propagating one-dimensional edge modes appears as a promising strategy towards an efficient phase-locking of the laser oscillation at the different sites. In this way, gain can be distributed over a large number of sites, while maintaining a globally stable single-mode coherent emission, which is very promising to realize high-power coherent sources. Experiments along these lines were reported shortly afterwards using photonic crystals under a strong magnetic field~\cite{Bahari2017} and arrays of coupled ring microcavities,~\cite{Bandres2018} followed by more recent valley-Hall quantum cascade~\cite{Zeng2020} and telecom-wavelength~\cite{Noh2020} lasers.

These experimental advances have stimulated an active theoretical research to characterize the peculiar properties of the novel devices.~\cite{Longhi2018,Malzard2018,Secli2019,Amelio2019} Whereas the experiments in Refs.~\onlinecite{Bahari2017,Bandres2018} have shown a clean single-mode emission from topolasers, the possibility of secondary instabilities as a result of the interplay of optical nonlinearities and slow carrier dynamics has been theoretically pointed out in Ref.~\onlinecite{Longhi2018}. A semiclassical study of the novel features introduced by the chirality of the lasing state was reported in Refs.~\onlinecite{Secli2017,Secli2019}. Extensive theoretical studies based on a stochastic approach have anticipated the robustness of the long-time coherence against static disorder by including quantum and thermal fluctuations into an idealized model of topolaser.~\cite{Amelio2019} Finally, the specific features of the weaker topologically protected but experimentally much less demanding topological lasing in valley Hall systems have been theoretically discussed in Refs.~\onlinecite{Gong2020,Zhong2020}.

In this work, we investigate the various mode-selection mechanisms that determine whether a topolaser device is going to lase in an edge or in a bulk state. Rather than dealing with the complex nonlinear dynamics of the lasing state,~\cite{Longhi2018,Secli2019,Amelio2019} we focus our attention on identifying the mode that is responsible for the first instability of the vacuum state. This is a common strategy in laser physics~\cite{Tureci2006} and typically provides a good intuition on the system behaviour not too far above threshold. For instance, if different modes of the laser resonator have different spatial profiles, a specific mode can be selected just by increasing its spatial overlap with the gain material. In the context of topolasing, a suitable spatial distribution of gain/losses was exploited in Ref.~\onlinecite{Zhao2018} to favour laser operation in protected zero mode of  a one-dimensional lattice. Theoretical work on the impact of nonlinearities on such spatial mode selection mechanisms and on possible transition to complex regimes with time-dependent power oscillations were reported in Ref.~\onlinecite{Malzard2018}. 

Here, we go beyond these works by including the additional spectral mode-selection mechanism coming from the frequency-dependence of gain. In its simplest formulation, spectral mode-selection allows to suppress competing cavity modes that are well separated in frequency by tuning a narrow-band gain material in the spectral vicinity of the desired mode. In particular, we take motivation from the recent topolaser experiment in Ref.~\onlinecite{Bahari2017} to investigate how a subtle combination of spectral and spatial mode-selection mechanisms can conspire to stabilize laser oscillation into a chiral edge state. As an important outcome of our analysis, we point out a possible mechanism for the still unexplained experimental observation~\cite{Bahari2017} of single-mode emission under a homogeneous pump with no need of concentrating pumping along the edge as it was instead done in other topolaser realizations, e.g., in Ref.~\onlinecite{Bandres2018}.

From a conceptual perspective, topological lasing under spatially homogeneous pumps is of special interest as it allows for a direct connection to the general concepts of non-Hermitian topology:~\cite{Ota2020,Bergholtz2021,Kawabata2019} in such a geometry, each region maintains in fact its (discrete) translational invariance. Bulk bands can thus be classified in terms of suitably generalized non-Hermitian topological invariants including the effect of gain and losses within the unit cell, and the value of the topological invariants can be then connected to the presence and the properties of edge states at the boundaries. 

The structure of the article is the following. In Sec.~\ref{sec:model} we introduce the physical system under investigation, namely a photonic Harper-Hofstadter lattice embedding population-inverted two-level atoms as gain medium, and we develop the theoretical model based on a Bloch-Harper-Hofstadter set of equations. In Sec.~\ref{sec:narrowband}, we show how the use of a narrow-band gain stabilizes the edge mode lasing even when the gain material is uniformly distributed across the whole system. In Sec.~\ref{sec:broadband} we show how a suitable combination of spectral and spatial selection mechanisms is able to stabilize the edge mode lasing under weak conditions on the gain lineshape and its spatial localization. The experimental implications of our results are discussed in Sec.~\ref{sec:discussion}. Conclusions are finally drawn in Sec.~\ref{sec:conclusions}. Additional details on the derivation of the theoretical models, on the topological lasing features, and on our spatial-spectral mode selection mechanism are given in the \suppmat{}.

\section{The theoretical framework}
\label{sec:model}

\begin{figure*}[!t]
    \centering
    \includegraphics[scale=0.38]{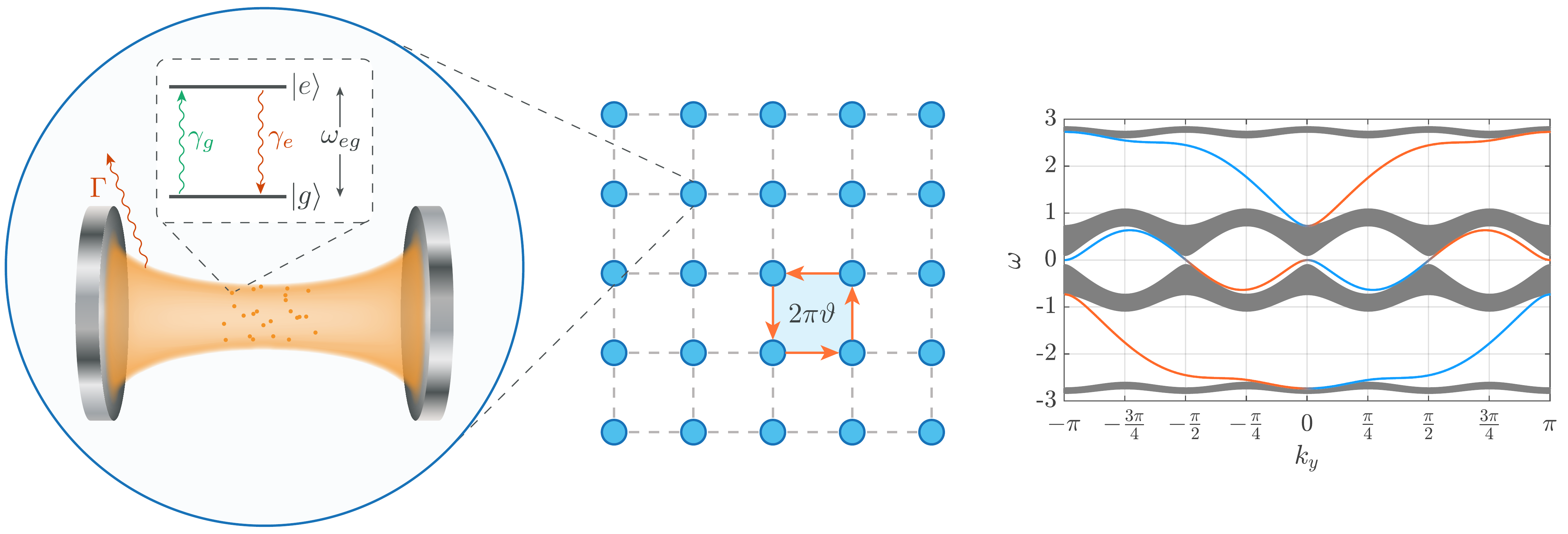}
    \caption{Left and central panels: scheme of a Harper-Hofstadter lattice consisting of an array of photonic cavities embedding two-level atoms (TLAs). The energy difference between the two atomic levels is $\omega_{eg}$; pumping of the atoms from $\Ket{g}$ to $\Ket{e}$ occurs at a rate $\gamma_g$, while their spontaneous decay from $\Ket{e}$ to $\Ket{g}$ occurs at a rate $\gamma_e$. All cavities decay at an equal loss rate $\Gamma$. The synthetic magnetic field is included as a non-trivial hopping phase: a photon that hops around a plaquette (orange arrows) picks up an extra phase $2\pi\vartheta$ due the synthetic magnetic flux, in our case $\vartheta = 1/4$. Right panel: plot of the energy dispersion for a $\vartheta = 1/4$ Harper-Hofstadter lattice with periodic boundary conditions along $y$ and $N_x=39$ sites with open boundary conditions along $x$. The gray bands correspond to bulk states. The colorful lines indicate the topologically protected edge states, red (blue) color indicating localization on the right (left) edge.}
    \label{fig:cavity_TLA}
\end{figure*}

In this Section we summarize the theoretical model used for our calculations. As an archetypal model, we consider a photonic Harper-Hofstadter lattice where optical gain is introduced  by including population-inverted two-level atoms at each site. For a complete derivation of the equations of this \emph{Bloch-Harper-Hofstadter} model, we refer the interested reader to Sec.~\suppref{S.1} of the \suppmat{}.

Harper-Hofstadter lattices were realized in integrated photonic devices by engineering the hopping links between neighboring microring resonators~\cite{Hafezi2013} and, in this form, were used in the topolaser experiment of Ref.~\onlinecite{Bandres2018}. Even though our study makes use of the Harper-Hofstadter model as a paradigmatic example of topological lattice, our conclusions extend to a wide variety of discrete or continuous topological photonics systems~\cite{Ozawa2019} and, in particular, help shining light on the photonic crystal experiment of Ref.~\onlinecite{Bahari2017}.

In the recent experimental implementations, gain is obtained by inserting  optically pumped quantum wells~\cite{Bahari2017,Bandres2018} or electrically driven quantum cascade heterostructures~\cite{Zeng2020} into a solid-state topological lattice. Here we will not dwell into the complexities of the microscopic physics of specific gain materials, but we will base our discussion on a simplest description in terms of population-inverted two-level atoms. In spite of its simplicity, this approach provides a reasonably accurate effective description of a wide range of actual media and, in particular, is able to correctly include the frequency-dependence of gain that is the focus of our analysis.

\subsection{The Bloch-Harper-Hofstadter model}
\label{sec:bloch_harper_hofstadter}

We consider a two-dimensional Harper-Hofstadter lattice~\cite{Hofstadter1976,Ozawa2019} where neighboring sites are connected through a hopping Hamiltonian with a non-trivial hopping phase. In the Landau gauge, this can be written as
\begin{multline}
    H_{\mathrm{bare}} = \hbar\omega_{\mathrm{cav}} \sum_{m,n} a_{m,n}^{\dagger}a_{m,n} \\ -J\sum_{m,n} \Big\lbrace a_{m,n}^{\dagger}a_{m+1,n} + e^{-i 2\pi\vartheta m}a_{m,n}^{\dagger}a_{m,n+1} + \text{h.c.} \Big\rbrace
    \label{eq:HH_tight_binding}
\end{multline}
where $a_{m,n}^{\dagger}$ ($a_{m,n}$) is the operator that creates (annihilates) a  photon at the $(m,n)$ site. All sites are assumed to have a bare photon frequency $\omega_{\mathrm{cav}}$, the real and positive parameter $J$ quantifies the hopping strength, and $\vartheta = 1/4$ is the flux per plaquette of the synthetic gauge field that is responsible for the topological properties. 

The topological features are easiest understood by considering a strip geometry with periodic boundary conditions along one direction ($y$) and open boundary conditions along the other ($x$). As it is shown in the right panel of Fig.~\ref{fig:cavity_TLA}, the energy dispersion in the first Brillouin zone $k_y \in [-\pi,\pi]$ shows four bulk bands: the central two bands touch at Dirac points, while the two external ones are separated by finite bandgaps, symmetrically located at positive and negative energies. The Chern numbers of the bands are, from bottom to top, $\mathcal{C} = -1,\,+2,\,-1$, where the two central bands have been considered as a single band in the calculation of the Chern number because of the degeneracy in the Dirac points. In agreement with these Chern numbers, each bandgap hosts one chiral edge mode and the edge modes in each topological gap have opposite chiralities, the one in the negative (positive) energy gap propagating in the counter-clockwise (clockwise) direction. For the strip geometry considered in this plot, two edge modes are present in each gap, unidirectionally propagating in opposite directions along $y$. Their topological protection from disorder stems from their spatial localization on opposite $x=1,N_x$ edges of the lattice, which suppresses back-scattering.

As it is sketched in Fig.~\ref{fig:cavity_TLA}, each site is modeled as a photonic resonator of frequency $\omega_{\mathrm{cav}}$ and decay rate $\Gamma$, and in which a photon is created (annihilated) by $a^{\dagger}$ ($a$). Each resonator is provided with a frequency-dependent gain medium, which is modeled here as a collection of $N$ population-inverted two-level atoms (TLAs). At each site, the dynamics of the atoms forming the gain medium is then described by the following Hamiltonian:
\begin{equation}
    H_{\mathrm{at},N}
    = \sum_{j=1}^N H_{\mathrm{at},1}^{(j)}
    = \sum_{j=1}^N \Bigg\lbrace \frac{1}{2}\hbar\omega_{eg}\sigma^z_j
    + \hbar g \Big( \sigma^+_j a + a^{\dagger} \sigma^-_j \Big) \Bigg\rbrace
    \label{eq:HH_cav}
\end{equation}
where $\omega_{eg} = \omega_e - \omega_g$ is the energy difference between the atomic levels, the light-atom coupling $g$ is assumed to be equal for all $j=1,\ldots, N$ atoms and, for each atom, $\sigma^+ = \Ket{e}\Bra{g}$ and $\sigma^- = (\sigma^+)^{\dagger} = \Ket{g}\Bra{e}$ are the raising and lowering operators between the ground $\Ket{g}$ and excited $\Ket{e}$ states. Analogously, the atomic population difference on each atom is quantified by the $\sigma^z = \Ket{e}\Bra{e} - \Ket{g}\Bra{g}$ operator. Each TLA is incoherently pumped from the ground to the excited state at a pumping rate $\gamma_{g}$, while the reverse spontaneous decay from $\Ket{e}$ to $\Ket{g}$ occurs at a rate $\gamma_e$. 

Under a mean-field approximation, we replace the photon field operators ${a}_{m,n}$ with their classical C-number expectation values $\alpha_{m,n} = \Braket{{a}_{m,n}}$ and we assume the atomic density matrix to have a factorized form. The one-atom Hamiltonian terms $H_{\mathrm{at},1}^{(j)} \equiv H_{\mathrm{at},1}$ and the one-atom density matrices $\rho_{\mathrm{at}}$ are then equal for all atoms at a given site and their dynamics is captured by a Lindblad master equation of the form
\begin{align}
    \frac{d\rho_{\mathrm{at}}}{dt} =
    &-\frac{i}{\hbar} [H_{\mathrm{at},1},\rho_{\mathrm{at}}] \nonumber \\
    &+ \sum_{s=e,g}\gamma_s\left( L_s\rho_{\mathrm{at}}L_s^\dagger -\frac{1}{2}\left\lbrace L_s^\dagger L_s, \rho_{\mathrm{at}}\right\rbrace \right)\,,
    \label{eq:Lindblad_at}
\end{align}
where the first term gives the coherent evolution induced by the atom-field dynamics in \eqref{eq:HH_cav} and $L_e = \Ket{g}\Bra{e} = \sigma^-$ and $L_g = \Ket{e}\Bra{g} = \sigma^+$ are the jump operators for the decay and pumping processes.

Projecting \eqref{eq:HH_cav} and \eqref{eq:Lindblad_at} onto the atomic ground and excited states then recovers the Bloch equations of the semiclassical theory of lasers.~\cite{Sargent1974,Scully1997} Together with the field dynamics determined by the hopping Hamiltonian \eqref{eq:HH_tight_binding}, these equations constitute the full set of equations of our Bloch-Harper-Hofstadter model. Measuring all energies and times in units of $J$ and $J^{-1}$ respectively, and setting the frequency zero at the empty cavity frequency $\omega_{\mathrm{cav}}$, these equations have the form:
\begin{equation}
    \begin{cases}
    \displaystyle \dot{\rho}_{ee}^{m,n} = \gamma_g \rho_{gg}^{m,n} - \gamma_{e}\rho_{ee}^{m,n} + i\Big( \alpha_{m,n}\rho_{ge}^{m,n} - \alpha^{*}_{m,n}\rho_{eg}^{m,n} \Big) \\
    \displaystyle \dot{\rho}_{gg}^{m,n} = \gamma_e \rho_{ee}^{m,n}- \gamma_{g}\rho_{gg}^{m,n} - i\Big( \alpha_{m,n}\rho_{ge}^{m,n} - \alpha^{*}_{m,n}\rho_{eg}^{m,n} \Big) \\
    \displaystyle \dot{\rho}_{eg}^{m,n} = -i\Big(\omega_{eg} - i\gamma\Big)\rho_{eg}^{m,n} - i\alpha_{m,n}\Big( \rho_{ee}^{m,n} - \rho_{gg}^{m,n} \Big) \\
    \begin{aligned}[c]%
	    \displaystyle \dot{\alpha}_{m,n} 
	    = &-\Gamma\alpha_{m,n} + iG\rho_{eg}^{m,n} + i\Big( \alpha_{m+1,n} + \alpha_{m-1,n} \\
	    &+ e^{-i 2\pi\vartheta m}\alpha_{m,n+1} + e^{+i 2\pi\vartheta m}\alpha_{m,n-1} \Big)
	\end{aligned}
    \end{cases}
    \label{eq:topolaser_TLA_equations}
\end{equation}
where $\rho_{ee}^{m,n}$ ($\rho_{gg}^{m,n}=1-\rho_{ee}^{m,n}$) is the average atomic population of the excited (ground) state of the atoms located at site $(m,n)$ and $\rho_{eg}^{m,n}$ is the corresponding coherence. In the following, we will assume for simplicity that no additional decay channel is acting on the atomic coherence $\rho_{eg}$ in addition to the unavoidable ones coming from pumping and decay, $\gamma=\gamma_{eg} = {(\gamma_g+\gamma_e)}/{2}$.

The efficiency of the gain process enters via the $G \doteqdot g^2 N$  coupling strength, proportional to the number of atoms $N$ per site and to the square of the elementary light-atom coupling $g$. Indicating with $V$ the effective field volume at each site, the light-atom coupling scales as usual as $V^{-1/2}$ (see Sec.~\suppref{S.1} of the \suppmat{}), which makes $G$ proportional to the atomic density $N/V$. If a real atomic gas is used as the gain material, the gain strength can be tuned by changing the density of the gas. In a solid state photonic crystal where the TLAs are used to model a more complex electronic dynamics in the material, the same effect can be achieved by varying the filling factor within the unit cell and/or the overlap of the Bloch mode with the gain material. To keep our analysis as simple as possible, in the following we will use $G$ as the  parameter controlling the strength of the gain in the different regions of space. This choice automatically includes the possibility of having different atomic densities in the different regions; compared to the pumping rate $\gamma_g$, it also allows to simplify the presentation by avoiding the complications due to the simultaneous dependence of several other parameters on $\gamma_g$.

The numerical results that we are going to present in the next Sections were obtained by numerically simulating the system evolution described by equations \eqref{eq:topolaser_TLA_equations} via a standard 4-th order Runge-Kutta integration scheme that provides direct access to time-dependent quantities. The steady-state values of observable are extracted by running the time-dependent simulations up to very long times. The lasing frequency and the power spectral density are obtained by Fourier transform of the late-time temporal evolution of the relevant field amplitudes.

\subsection{Frequency-dependent gain}

The peculiar features of the frequency-dependent gain can be simply understood by looking at the laser operation in a single site geometry. Indicating with $\omega_L$ the lasing frequency, the steady-state is determined by the late time behaviour of the solution to the Bloch equations \eqref{eq:topolaser_TLA_equations} in the single-site case.

In this limit, the atomic populations tend a constant value, while both the coherence $\rho_{eg}$ and the field amplitude $\alpha$ keep oscillating at frequency $\omega_L$. Explicit expressions for the steady-state atomic quantities are given in the \suppmat{}. The steady-state field amplitude is instead $\alpha(t) = \tilde{\alpha}\,e^{-i\omega_L t}$ with the amplitude $\tilde{\alpha}$ satisfying the gain/loss balance equation
\begin{equation}
    \frac{P}{1+\beta|\tilde{\alpha}|^2} = \Gamma
    \label{eq:alpha_steady_state}
\end{equation}
with the effective pump strength
\begin{equation}
    P
    = G
    \left(\frac{\gamma}{(\omega_{eg}-\omega_L)^2 + \gamma^2}\right)
    \left(\frac{\gamma_g-\gamma_e}{\gamma_g+\gamma_e}\right)
    \label{eq:P_effective}
\end{equation}
and the saturation coefficient
\begin{equation}
    \beta 
    = 2\frac{g^2}{\gamma_{eg}}
    \left(\frac{\gamma}{(\omega_{eg}-\omega_L)^2 + \gamma^2}\right).
\end{equation}
Equation \eqref{eq:alpha_steady_state} is formally analogous to the one of a broadband saturable gain considered in Refs.~\onlinecite{Harari2018,Secli2019}, with the key difference that the parameters $P,\beta$ are here frequency-dependent. In particular the effective pump strength $P$ involves a Lorentzian factor ${\gamma}/{[(\omega_{eg}-\omega_L)^2 + \gamma^2]}$ accounting for the non-trivial gain spectrum: this is centered at $\omega_{eg}$ and has a HWHM set by the atomic decoherence rate $\gamma$. 

This frequency-dependent gain directly reflects into an analogous dependence of the laser threshold. In a single-site geometry this is immediately obtained from Eqs.~\eqref{eq:alpha_steady_state} and \eqref{eq:P_effective} as the lowest value of $G$ for which (unsaturated) gain exceed losses $P \geq \Gamma$. This leads to the threshold condition
\begin{equation}
    G > G_{\mathrm{res},0}\left[1 + \left(\frac{\omega_L-\omega_{eg}}{\gamma}\right)^2\right],
    \label{eq:LT_condition}
\end{equation}
where
\begin{equation}
    G_{\mathrm{res},0} \doteqdot \frac{2\gamma_{eg}\gamma\Gamma}{\gamma_g-\gamma_e}
    \label{eq:G_res_0_def}
\end{equation}
is the single-cavity lasing threshold exactly on resonance, that is for $\omega_{L} = \omega_{eg}$. As expected, the threshold is minimum when laser operation occurs on resonance with the pupolation-inverted atoms. Then, it increases quadratically with the detuning $\omega_{L}-\omega_{eg}$: the faster the atomic decoherence rate $\gamma$, the weaker this increase. In the following of the work, we will exploit this frequency-dependence of the threshold as a way to select the desired mode for lasing.

\section{Narrowband Gain}
\label{sec:narrowband}

\begin{figure*}[!t]
    \centering
    \begin{minipage}[c]{0.5\textwidth}
        \hspace*{-12pt}
        \hbox{
            \includegraphics[scale=0.5,valign=b]{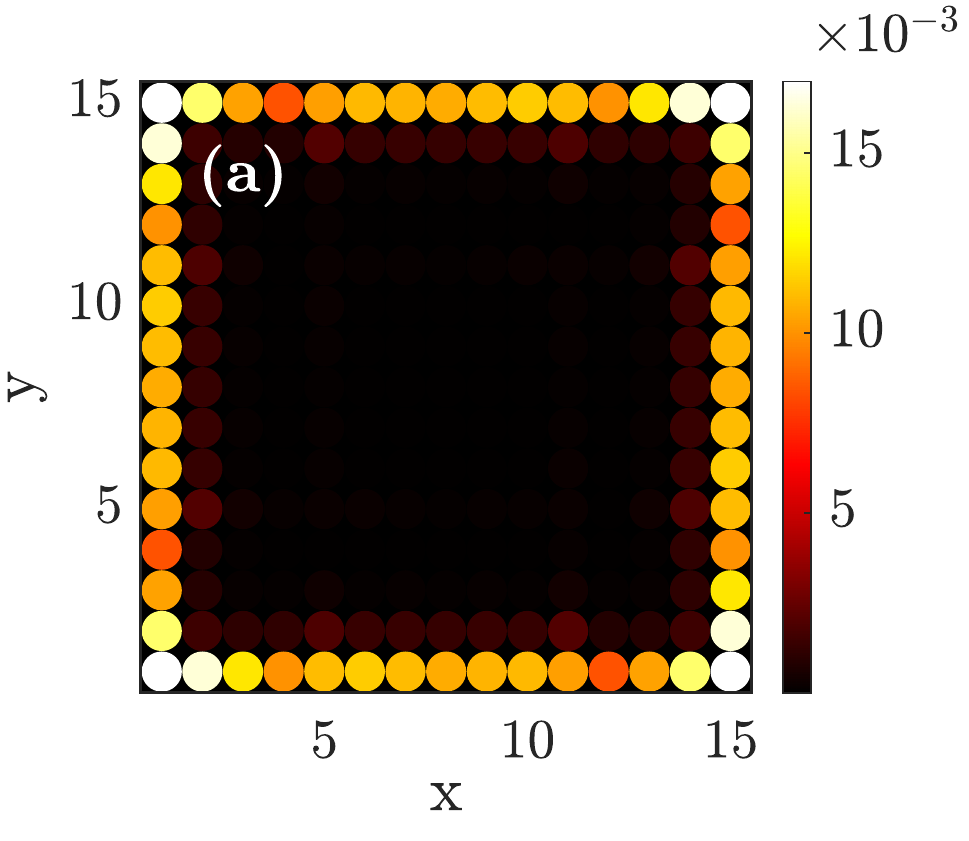}
            \includegraphics[scale=0.5,valign=b]{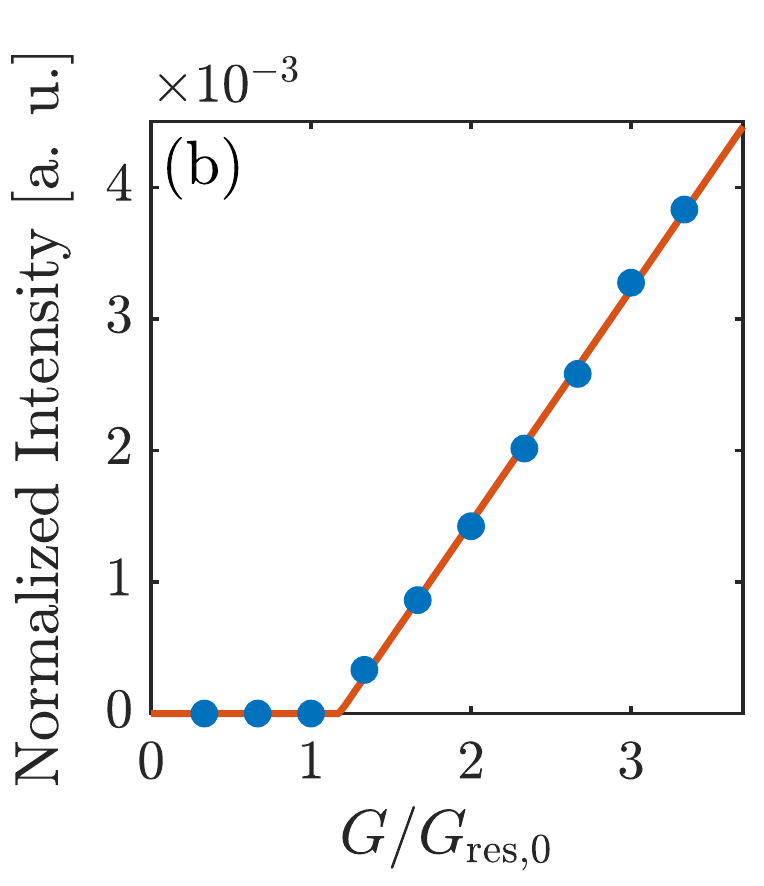}
        }
        \hspace*{-12pt}\includegraphics[scale=0.5,valign=b]{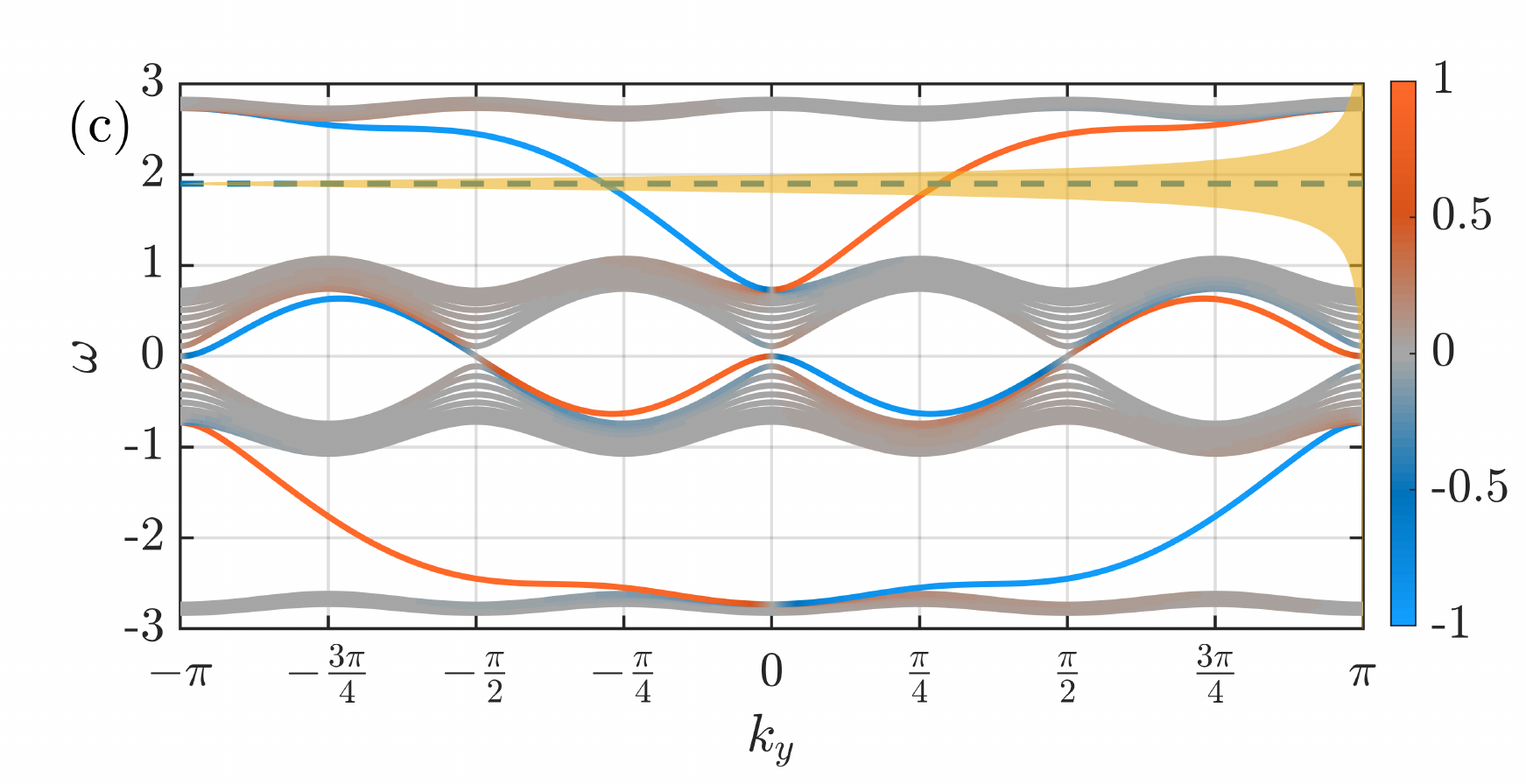}
    \end{minipage}
    \begin{minipage}[c]{0.48\textwidth}
        \hspace*{-12pt}
        \includegraphics[scale=0.5,valign=b]{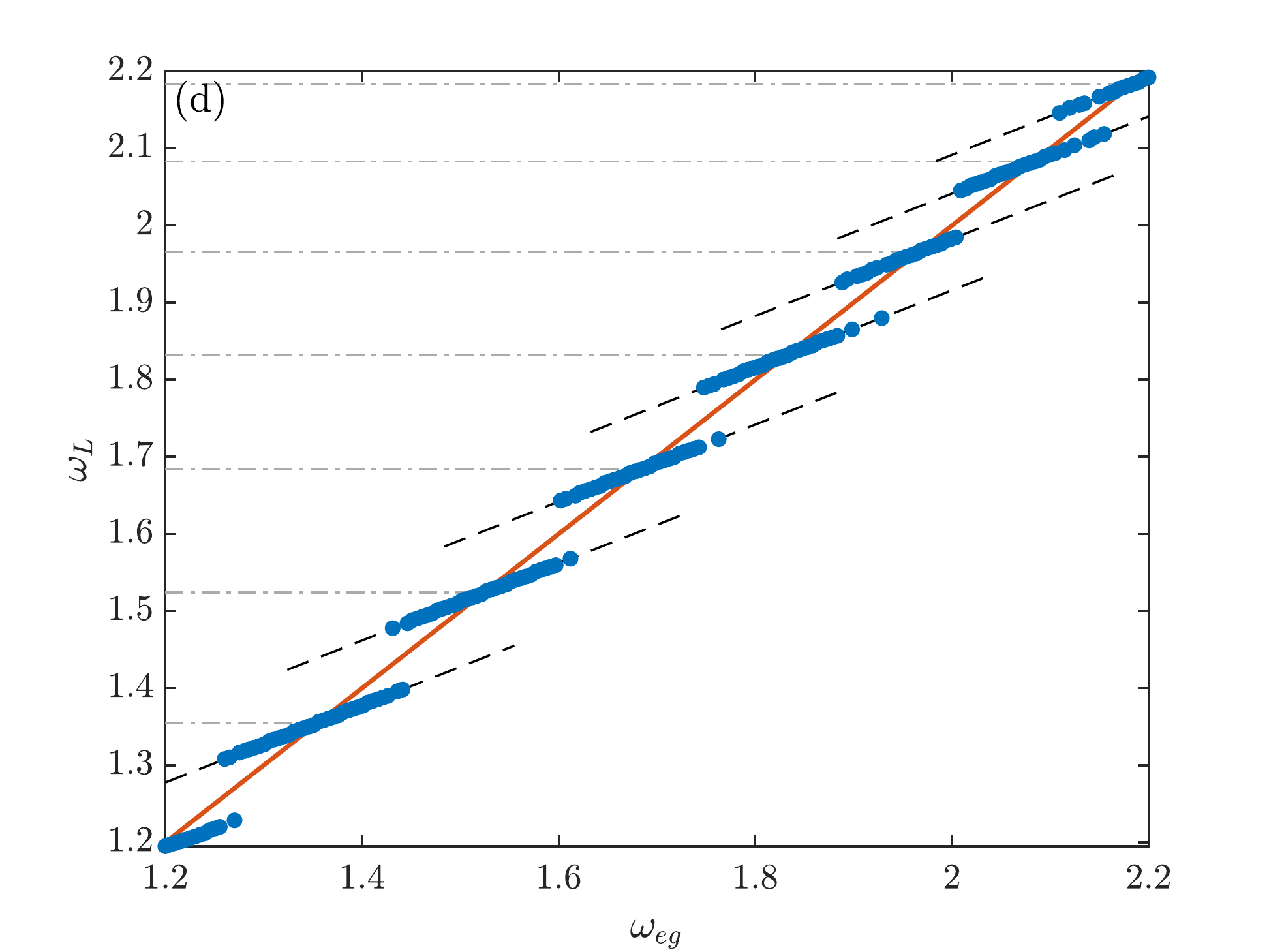}
    \end{minipage}
    \caption{Panel (a): steady state emission of a $15 \times 15$ lattice, uniformly pumped on all sites with a narrowband gain of strength $G/G_{\mathrm{res},0} = 3$. The atomic transition frequency is $\omega_{eg} = 1.9$. For these parameters, lasing is found to occur at $\omega = 1.87$.
    Panel (b): spatially averaged emitted intensity as a function of the gain strength $G$, showing the typical linear behavior after the lasing threshold.
    Panel (c): band structure of a Harper-Hofstadter lattice with PBCs in the $y$-direction but finite in the $x$-direction $N_x = 39$. The colorscale from blue to red quantifies the localization of each mode from left to right in the $x$-direction, while the dashed line indicates the atomic resonance $\omega_{eg}$. The narrowband gain used for the simulations in the panels above is represented as a Lorentzian in the frequency domain, centered at $\omega_{eg}$ and with a FWHM $2\gamma = 0.2$, i.e.\@ roughly $13\%$ of the topological bandgap.
    Panel (d): numerically observed lasing frequency (blue dots) as a function of the atomic resonance frequency $\omega_{eg}$. Each slant step is fitted with a black, dashed line with equation $\omega_L = \frac{1}{2}\omega_{eg} + b$. The light grey dash-dot lines mark the lasing frequency values for which $\omega_L = \omega_{eg} = \omega_0$ (see \eqref{eq:MP_lattice}), while the red line represents the approximated curve $\omega_{L} = \omega_{eg}$. The simulations have been performed by numerically integrating the Bloch-Harper-Hofstadter motion equations \eqref{eq:topolaser_TLA_equations} up to a time $T=500\,\Gamma^{-1}$; the other parameters are the same as in the left panels. All frequencies are measured in units of the hopping $J$ and the zero is at the bare cavity frequency $\omega_{\mathrm{cav}}$. The lasing frequency is extracted from the power spectral density of the emission that is obtained by a temporal Fourier transform of the light field amplitude in the latest $\Delta T=150\,\Gamma^{-1}$ of the evolution.}
    \label{fig:narrowband}
\end{figure*}


In the previous Section we have restricted our attention to the single-site case. This provides us the conceptual building blocks to understand laser operation in a topological lattice. As a first step in this direction, in this Section we consider the simplest case where the narrowband gain spectrum is concentrated within a topological gap. In contrast to the chaotic multi-mode emission found in Ref.~\onlinecite{Secli2019} for the extreme broadband gain, here we show that such narrowband gain can lead to a stable topological lasing even under a spatially uniform pumping. While such a narrowband gain might not be the technologically simplest option for practical devices, a detailed discussion of its features is an interesting first step to validate our Bloch-Harper-Hofstadter model and understand its behaviour in the different regimes. In addition to that, because of mode-pulling effects, a interesting non-trivial relation is found between the lasing frequency and the bare frequencies of the discrete set of edge modes.

\subsection{Single-mode topological laser emission}

This narrowband gain configuration can be obtained by considering the Bloch-Harper-Hofstadter model introduced in Sec.~\ref{sec:bloch_harper_hofstadter} and tuning the atomic frequency $\omega_{eg}$ in the middle of the topological bandgap with a gain linewidth $\gamma$ much smaller than the gap width, as sketched in Fig.~\ref{fig:narrowband}(c). In this way, the frequency-dependence of gain strongly increases the effective threshold for laser operation in the off-resonant bulk band states while the one for edge state lasing remains almost unaffected.

Laser operation in this regime is illustrated in Fig.~\ref{fig:narrowband}(a--b). Emission into the edge state is stable and monochromatic and remains so up to high pump strengths well above the laser threshold.~\cite{Secli2019a} Quite interestingly, such monochromatic single-mode emission is not restricted to small lattice sizes where a single eigenstate --- classified by $k_y$ for the strip geometry of Fig.~\ref{fig:narrowband}(c), or by the winding number around the lattice for the geometry of Fig.~\ref{fig:narrowband}(a) --- falls within the gain bandwidth: as it was pointed out in Ref.~\onlinecite{Secli2019} for the extreme broadband gain case, the high spatial overlap of different edge states provides in fact a very efficient mode competition mechanism~\cite{Sargent1974} eventually suppressing simultaneous laser operation in multiple modes. The dynamical stability of the single mode emission was confirmed by the Bogoliubov analysis in Ref.~\onlinecite{Loirette-Pelous2021}. A further illustration of the dynamics of this mode-competition process over time is provided in Sec.~\suppref{S.3} of the \suppmat{} where we display a time-frequency representation of the lasing process. As usual, the choice of the specific lasing mode is stochastically determined at each instance of lasing by the initial conditions and the noise. Still, for sufficiently narrow-band gain, the resulting probability distribution for lasing in different modes will be very peaked on the most likely mode. 

Of course, this monochromatic emission only holds up to moderate pump strengths at which only the quasi-resonant edge modes experience an effective gain. At very high pump strength also the bulk modes go above threshold and the dynamics recovers the chaotic behaviour found in Ref.~\onlinecite{Secli2019} for a broadband gain distributed in the whole system.

\subsection{Mode-pulling effects}

In contrast to the broadband gain case where the laser frequency $\omega_L$ is typically locked to the bare mode frequency $\omega_0$, for a narrowband gain  a sizable \emph{mode pulling} effect can occur on the laser frequency.~\cite{Scully1997} The lasing frequency results then from a weighted average of the atomic resonance $\omega_{eg}$ and the bare mode frequency $\omega_0$ via the mode pulling formula
\begin{equation}
    \omega_{L} = \frac{\omega_{0} + \mathcal{S}\omega_{eg}}{1+\mathcal{S}}\,,
    \label{eq:MP_lattice}
\end{equation}
where $\mathcal{S} = {\Gamma}/{\gamma}$ is the so-called the \emph{stabilization factor}. When $\mathcal{S} \ll 1$ mode pulling effects are negligible and $\omega_L = \omega_0$, where $\omega_0$ is the frequency of the corresponding lattice mode selected by the lasing process. For equal $\Gamma = \gamma$, the stabilization factor is $\mathcal{S}=1$ and the mode pulling effect becomes a simple average. Physically, this mode-pulling effect can be understood as the result of the refractive index change that is naturally associated to the gain via Kramers-Kronig causality relations: as usual, narrow resonances are responsible for quantitatively larger changes of the refractive index in their spectral neighborhood.

Let us explore the impact of this effect in our case of topological laser operation for a narrowband gain centered inside the topological bandgap, thus perfectly overlapping with the edge state dispersion. The blue dots in Fig.~\ref{fig:narrowband}(d) show the numerical prediction for the steady-state lasing frequency as a function of the atomic resonance position $\omega_{eg}$. This plot illustrates an interesting interplay of mode pulling with the intrinsic discreteness of the edge state.~\cite{Secli2019} Given the finite size of the system, the edge state consists in fact of a sequence of discrete states classified by the winding number around the perimeter of our square sample. In panel (d), the frequencies of such state are indicated by the horizontal dash-dotted lines.

When mode pulling effects are negligible, for instance in a broadband gain case, the system lases at the frequency of the edge mode that is closest to the resonance $\omega_{eg}$ for which gain is strongest. In our plot, this would correspond to a staircase of flat steps separated by a spacing $\Delta\omega = {2\pi v_g}/{L}$ determined by the overall length $L$ of the system edge and the group velocity $v_g$ of the edge mode. 

For our narrowband gain, laser operation still occurs in the discrete mode that is closest to $\omega_{eg}$, but mode pulling effects make the steps to have a finite slope instead of being flat. Inserting the frequency of the selected edge state as the cavity frequency $\omega_0$ in \eqref{eq:MP_lattice} predicts a value $\frac{\mathcal{S}}{1+\mathcal{S}}$ for the slope. In the figure we have taken $\gamma=\Gamma$, so the stabilization factor is $\mathcal{S}=1$ and the expected slope is $1/2$. This value (black dashed lines) is in perfect agreement with the numerical findings.

\section{Broadband Gain}
\label{sec:broadband}

\afterpage{\onecolumngrid\par
\begin{figure}[p]  
    \centering
    \begin{minipage}[b]{0.50\textwidth}
        \hspace*{-36pt}
        \includegraphics[scale=0.5]{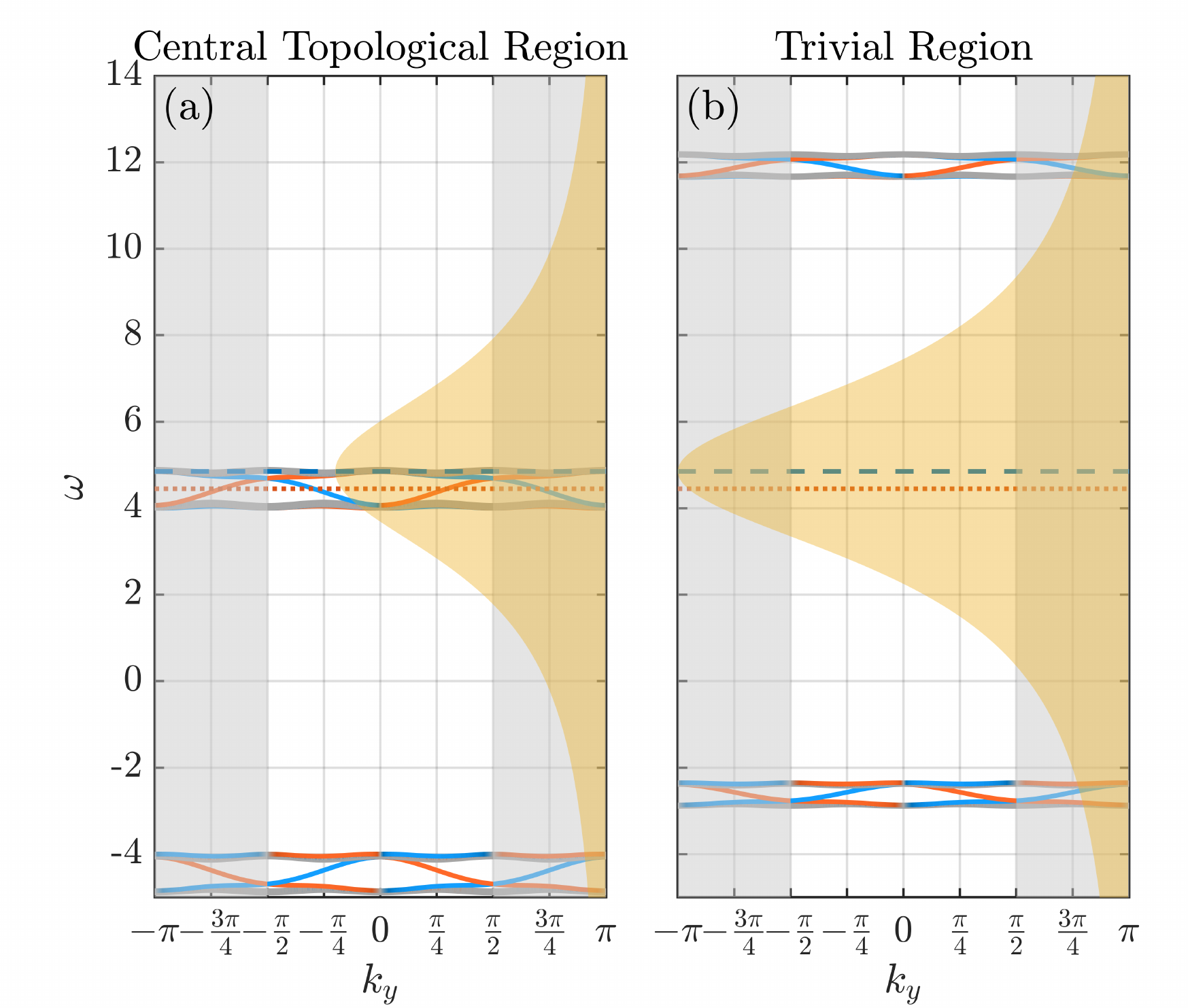}\\ 
        \hspace*{-27pt}
        \includegraphics[scale=0.5]{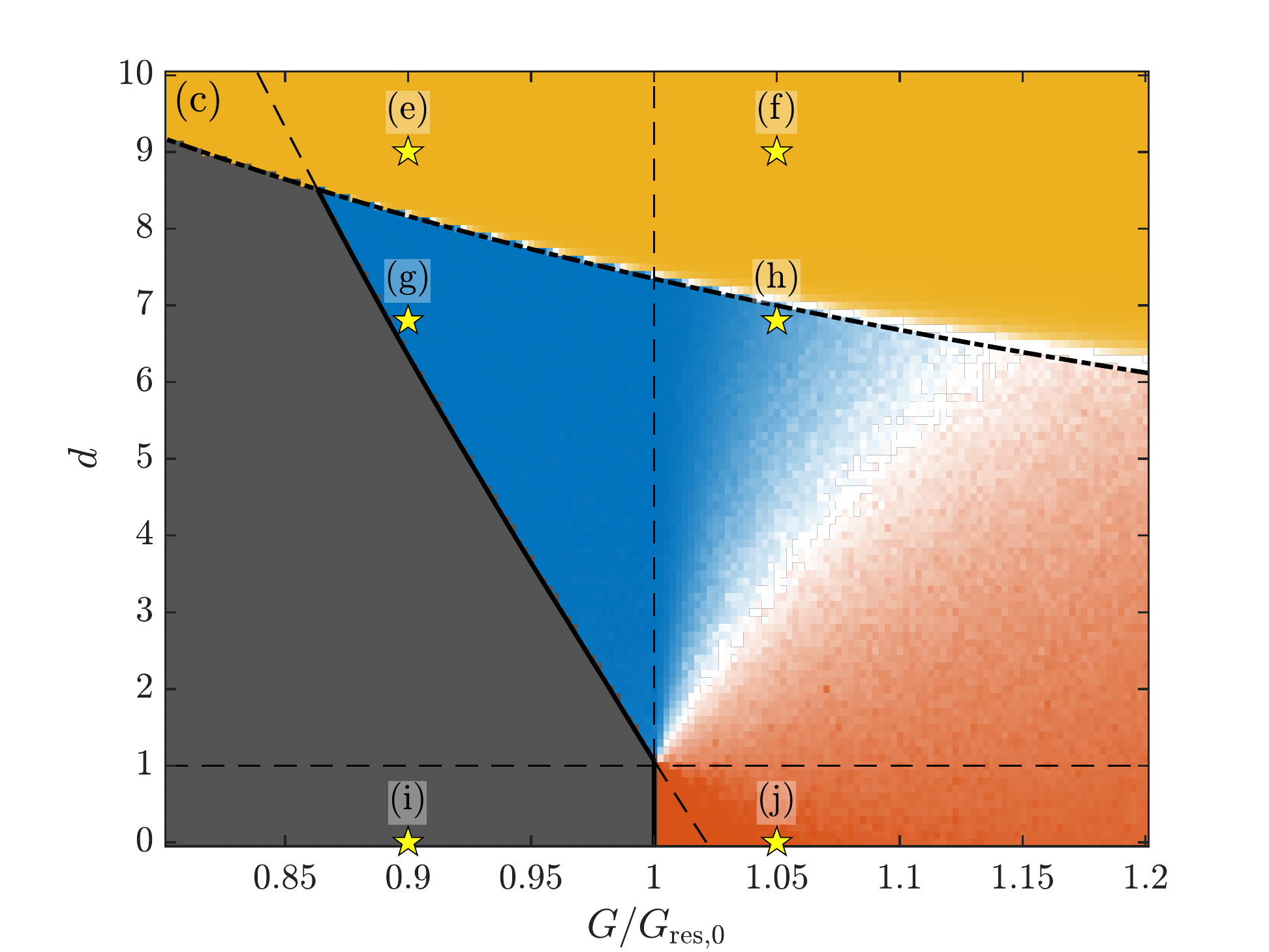}
    \end{minipage}
    \begin{minipage}[b]{0.48\textwidth}
        \hspace*{-27pt}
        \includegraphics[scale=0.5]{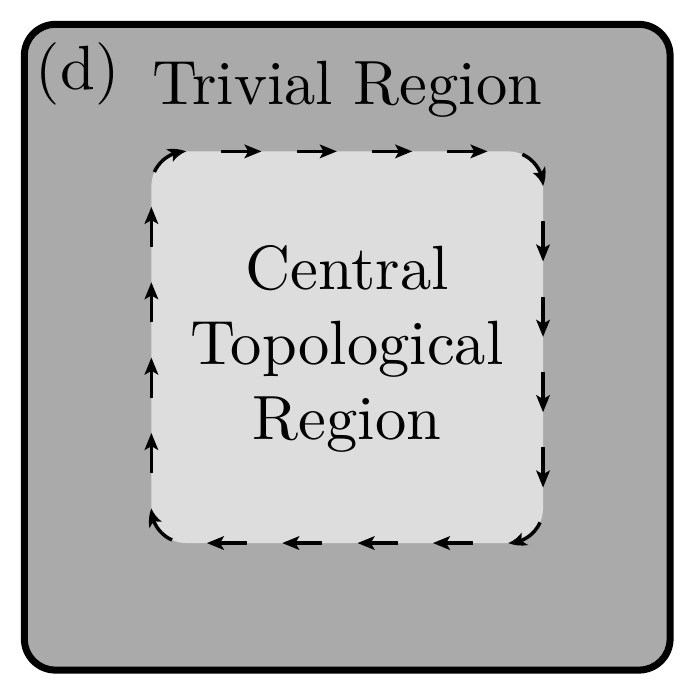}\\
        \hspace*{-27pt}
        \includegraphics[scale=0.5]{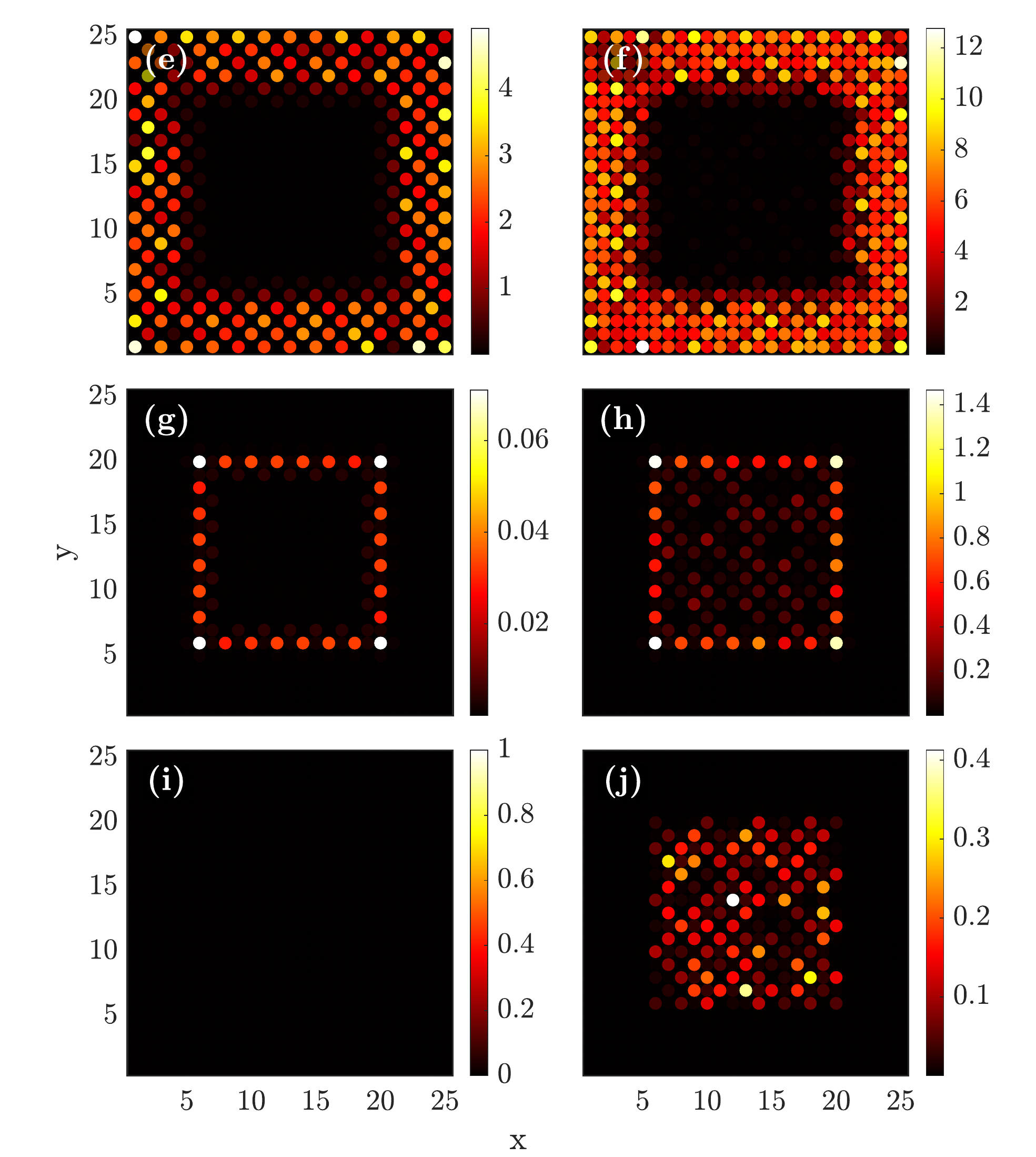}
    \end{minipage}
    \caption{Result of simulations performed on a $25\times25$ lattice with a $5$-sites thick surrounding region.
    Panels (a)--(b): band structure for the proposed broadband scheme, with a geometry schematically depicted in panel (d). The central region (panel (a)) is a bipartite $\vartheta=1/4$ Harper-Hofstadter lattice with checkerboard detuning $\Delta = 4.0$, while the surrounding region (panel (b)) has the same geometry with a larger checkerboard detuning $\Delta_{\mathrm{trivial}} = 7.0$ and a global detuning $\omega_{\mathrm{trivial}} = 4.65$ (red dotted line). The gain spectrum is centered at $\omega_{eg} = 4.85$ (blue dashed line) and has a FWHM linewidth $2\gamma = 5.2$ (yellow shading). The relative strength of gain on the two sides depends on $d=G_{\mathrm{trivial}}/G$: for the specific case illustrated in the figure, the taller gain spectrum in panel (b) refers to a $d>1$ case of stronger gain in the trivial region. The grey shaded areas indicate $k_y$-vectors outside the reduced Brillouin zone. For these parameters, the gain linewidth is around $8.2$ larger than the width of the topological band gap of the central region and around $37\%$ of the width of the trivial band gap in the surrounding region. It is around $63$ times wider than the one considered in Fig.~\ref{fig:narrowband}. 
    Panel (c): phase diagram of the different lasing regimes as a function of the overall pumping strength $G/G_{\mathrm{res},0}$ and of the relative effective density of gain material of the surrounding region $d = G_{\mathrm{trivial}}/G$ for the same lattice parameters used in the dispersion plots shown in the top left panels. A grey color indicates no lasing; a blue color indicates lasing from the topological edge mode; a red color indicates lasing from the non-topological portion of the central region; a yellow color indicates lasing from the surrounding region. Fading to white indicates the coexistence of multiple phases. The thin dashed black lines indicate the $G/G_{\mathrm{res},0} = 1$ and $d = 1$ values. The solid and dot-dashed black transition lines between different phases are analytically predicted via the prescriptions in Sec.~\suppref{S.5} of the \suppmat{}.
    The six yellow stars have a one-to-one correspondence with the six panels (e)--(j) presented on the right, showing sample snapshots of the real-space emitted intensity at the end of the integration time. The first, second and third row from the top are for decreasing values of $d = 9.0, \, 6.8, \, 0$ respectively. The left and right columns are for increasing $G/G_{\mathrm{res},0} = 0.90, \, 1.05$, respectively below and above the single-site resonant lasing threshold.
    All simulations have been performed by numerically integrating the Bloch-Harper-Hofstadter motion equations \eqref{eq:topolaser_TLA_equations} up to a time $T = 10^4\,\Gamma^{-1}$.}
    \label{fig:broadband_results}
\end{figure}
\clearpage\twocolumngrid}


In the previous Section, we have seen an efficient scheme to stabilize topolaser operation with a uniformly distributed gain, by spectrally concentrating the gain spectrum in the topological gap. While conceptually interesting, this scheme is hardly useful in practical semiconductor systems, where the gain linewidth is typically comparable if not larger than the width of the topological band gaps so that an efficient spectral selection of the edge mode from the neighboring bulk modes is hardly obtained.

In this Section we will explore a more sophisticated scheme that is able to stabilize topological lasing in a much wider range of parameters of potential technological relevance. The configuration we consider is inspired by the photonic crystal experiment in Ref.~\onlinecite{Bahari2017} where the topological band gap is orders of magnitude narrower that the gain linewidth. One of the aims of our work is to provide theoretical insight into the observed single-mode topolaser emission of this experiment. 

A sketch of the configuration under examination is shown in Fig.~\ref{fig:broadband_results}(d). As in the experiment,~\cite{Bahari2017} we consider a central region, which has a narrow topological gap, surrounded by a region with a much wider and topologically trivial gap. Chiral boundary modes are localized at the interface between the two regions. We also assume that the gain bandwidth is much larger than the narrow topological gap but comparable to the large trivial gap, as in the experiment, and that the gain is stronger in the trivial region than in the topological region. As a consequence, even though the gain material is pumped in a globally spatially uniform fashion we can expect clear topolaser operation in the edge states which partially penetrate into the trivial region with stronger gain, while lasing into the bulk states of the trivial region is suppressed by their detuning from the gain bandwidth.  In this Section, we explain in detail how this idea works.

We first explain how we model a narrow and isolated topological gap in the central region. We start from the $\vartheta=1/4$ Harper-Hofstadter lattice, which contains multiple topological band gaps with topological invariants adding up to zero. We want to isolate one topological band gap from these multiple gaps. To this end, we add a checkerboard-shaped on-site frequency detuning $\pm \Delta$: the frequencies of the $(m,n)$ sites are thus alternated and equal to $\omega_{\mathrm{cav}} + \Delta\cdot(-1)^{m+n}$. The photonic bands of such a \emph{bipartite} Harper-Hofstadter model are shown in panel (a) of Fig.~\ref{fig:broadband_results}: because of the checkerboard detuning, the Brillouin zone is reduced to $k_y \in \left[-\frac{\pi}{2},\frac{\pi}{2}\right]$ and the Dirac touching points between the middle two bands open into a trivial gap with size $\sim 2\Delta$. In agreement with the sequence $-1,+1,+1,-1$ of Chern numbers of the different bands, the two (small) gaps between the lower two bands and the upper two bands maintain their topological nature visible in the corresponding edge states. In what follows, we focus on the topological band gap of the two upper bands; the gain spectrum is centered around the frequency of the two upper bands. The two lower bands are, instead, off resonant, and are not relevant in the laser operation and the discussion below.

Next, we explain how we prepare the surrounding region with a wide trivial gap. We again start from the $\vartheta=1/4$ Harper-Hofstadter model and add a checkerboard-shaped detuning $\Delta_{\mathrm{trivial}}$, which is larger than $\pm \Delta$ in the topological region. We add a global shift of all site frequencies by $\omega_{\mathrm{trivial}}$ so that the large topologically trivial gap between the middle two bands is centered around the two upper bands of the topological region. The corresponding photonic bands are shown in panel (b). Although the two upper and two lower bands have narrow topological gaps, they are pushed away by the large $\Delta_\mathrm{trivial}$, and thus we can focus on the effect of the wide trivial gap between the middle two bands. We call this surrounding region a ``trivial'' region in this sense. The gain spectrum, which is indicated by the yellow shading in panels (a)--(b), is centered at the middle of the wide trivial gap and completely encompasses the topological gap in the central region.

The gain strength in the surrounding trivial region, $G_{\mathrm{trivial}}$, can be reinforced either by locally increasing the pumping strength or, alternatively, by keeping a spatially uniform pumping but increasing the density of gain material with respect to the central region, as discussed in Sec.~\ref{sec:bloch_harper_hofstadter}. Focusing on this latter case, which appears relevant for the experiment in Ref.~\onlinecite{Bahari2017}, we can write $G_{\mathrm{trivial}} = G \cdot d$, where $d$ can be interpreted as the effective density of gain material in the surrounding region relative to the central region, and treat $G/G_{\mathrm{res},0}$ as a global measure of the uniformly distributed pumping strength in units of the single-resonator resonant threshold.

The results of the numerical simulations are summarized in Fig.~\ref{fig:broadband_results}, where we show a phase diagram of the different regimes of laser operation as a function of the relative effective density of gain material in the surrounding region $d=G_{\mathrm{trivial}}/G$ and of the pumping strength in the central topological region in units of the resonant, single-site threshold, $G/G_{\mathrm{res},0}$.

When the surrounding trivial region is purely passive and does not display any gain ($d=0$, Fig.~\ref{fig:broadband_results}(i) and (j)), the system is almost equivalent to a bipartite $15\times15$ lattice without the surrounding region. We therefore expect the system to only lase above the resonant single-site lasing threshold, $G/G_{\mathrm{res},0} = 1$, as shown by the red region at the bottom of the phase diagram. Since the gain is effectively broadband with respect to the upper pair of photonic bands in the central region, both bulk and boundary modes equally participate in the lasing process, forbidding a stable topological laser operation (panel (j)). Compared to the bulk states, boundary states are even slightly disfavoured by the worse spectral overlap with the gain spectrum and by their evanescent tail that penetrates into the surrounding trivial region and reduces the spatial overlap with the gain region.

We can induce lasing from the topological edge modes at the boundary by making the gain in the surrounding trivial region to be stronger than the one in the central topological region, that is $d>1$. In this case, a region appears in the parameter space where the system displays a monochromatic topological laser behaviour (panel (g)). Thanks to their evanescent tail overlapping with the stronger amplifying surrounding region, the effective threshold of the topological boundary modes is in fact pushed well below the one $G/G_{\mathrm{res},0}=1$ of the bulk modes (thick solid black line in the phase diagram), opening a window where only these modes can lase (blue region). In this $G/G_{\mathrm{res},0}<1$ regime, the monochromaticity of the topolaser emission is ensured by the same mode competition effects pointed out in Ref.~\onlinecite{Secli2019} and reviewed in Sec.~\ref{sec:narrowband}: since only topological edge modes experience a sufficient gain to lase and since these modes spatially share the same active medium, steady-state lasing ends up being concentrated in one of them only, thus making the  emission monochromatic. A more detailed time-frequency analysis in support of this conclusion is reported in Sec.~\suppref{S.3} of the \suppmat{}. 

In order to better quantify the efficiency of our combined mode-selection scheme, we have investigated the constraints on the gain linewidth $\gamma$ in order to have pure topolaser operation into the edge state. Quite interestingly, topolasing turns out to be robust as long as the effective gain experienced by the bulk bands in the trivial region remains below threshold. As it can be inferred from the discussion in Sec.~\suppref{S.5} of the \suppmat{}, for sufficiently large $\Delta_{\mathrm{trivial}}$ the upper bound on $\gamma$ involves $\gamma/\Delta_{\mathrm{trivial}}$ only. In particular, no restriction applies to the ratio of $\gamma$ to the topological gap width. In the experiment of Ref.~\onlinecite{Bahari2017}, the analog of $\Delta_{\mathrm{trivial}}$ is way larger than the topological band-gap, which releases any constraint on the gain bandwidth compared to the topological gap and allows this latter to be arbitrarily small. 

Note that this spectral structure is characteristic of the photonic crystal platform of Ref.~\onlinecite{Bahari2017} and is different from the typical one of the ring-resonator-based platforms considered in Ref.~\onlinecite{Bandres2018}. Here, additional copies of the band structure are in fact present with a spacing set by the (relatively small) free spectral range of the single ring resonators and our mode-selection mechanism is not applicable in a straightforward way.

The situation is of course very different in the $G/G_{\mathrm{res},0}>1$ case, when also the bulk modes of the topological region go above threshold. Since these modes have a reduced spatial overlap among them and with the edge mode, mode competition is no longer effective in ensuring a monochromatic emission and the latter acquires a complex multi-mode character (blue-to-white-to-red region). Still, thanks to the stronger gain of the surrounding trivial region, the intensity of boundary mode lasing can remain significantly stronger than the one of the central bulk modes even at values of $G$ above the single-cavity lasing threshold (panel (h)).

For even higher values of the surrounding density $d$ above the dash-dotted black line in the phase diagram, we reach a point where the spectral selection is no longer sufficient to suppress bulk lasing in the surrounding region and topological lasing is no longer possible. In this phase (yellow area in the phase diagram), the much stronger gain of the surrounding region makes the laser emission to be concentrated in this region (panels (e) and (f)).

A quantitative analytical discussion of the location of the transition lines in the phase diagram is given in Sec.~\suppref{S.5} of the \suppmat{}. As expected, the area of the topological lasing region in parameter space can be increased by either increasing the trivial bandgap in the surrounding region or by using a narrower gain spectrum. This trend is confirmed by the additional numerical simulations with different values of the parameters that are shown in Sec.~\suppref{S.4} of the \suppmat{}.

\section{Discussion}
\label{sec:discussion}

In the previous Sections we have concentrated our attention on a Harper-Hofstadter model which provides a relatively straightforward insight into the basic effects, but our conclusions extend to any combination of lattices with suitable spectral and topological properties. In particular, we expect that our physical conclusions extend even outside the tight-binding approximation that has been made in all theoretical studies of topological lasing so far.

As a most intriguing example, the results of our calculations are compatible with some key observations of the pioneering experiment in Ref.~\onlinecite{Bahari2017} that, to the best of our knowledge, remain so far unexplained. In particular, topological lasing was observed in this experiment without the need to concentrate gain along the edge separating the topological and trivial regions as it was instead the case in other experiments.~\cite{Bandres2018,Zeng2020} A key difference between the devices used in these works consists in that the topological system used in Ref.~\onlinecite{Bandres2018} is surrounded by empty space, while in Ref.~\onlinecite{Bahari2017} the central topological system is surrounded by a topologically trivial region where the field can penetrate with a significant evanescent tail. Most importantly for our purposes, the outer region displays a larger filling factor of the unit cell (compare Figs.~2A and 2B in Ref.~\onlinecite{Bahari2017}). For an equal level of optical pumping, we can thus reasonably expect the gain to be stronger in the outer region, which corresponds to $d>1$ in our model. As a result, the overlap of the edge state with this stronger amplifying region favours topological lasing with respect to bulk lasing in the central region. At the same time, the much wider extension of the trivial photonic band gap of the outer region forbids laser operation in the outer region thanks to the natural frequency-dependence of gain in the used semiconductor quantum well material. 

While these arguments provide a suggestive interpretation of experimental observations, they are of course not yet completely sufficient to rule out other possible explanations. For instance, in analogy to the arguments put forward in Ref.~\onlinecite{Noh2020} for a different geometry, another potentially relevant mechanism for stabilizing the edge mode lasing could originate from the weaker losses of the edge mode compared to the ones of bulk modes.~\cite{Kante2021inpreparation} In the specific system of Ref.~\onlinecite{Bahari2017}, reduced radiative losses may in fact originate from the evanescent tail in the outer trivial region where bulk modes in the vicinity of the trivial gap are below the light cone. In our model theory, the reduced radiative losses of the trivial region could be explicitly included via a reduced $\Gamma$ of the outer sites, but we expect their effect to be similar to the one of the increased gain $G_{\mathrm{trivial}}$ considered in our calculations. On this basis we are confident that the qualitative conclusions of our theory directly apply to the experiment. However, a firm and definitive unraveling of these questions requires accurate experimental measurements and comprehensive microscopic calculations of the band structure and of the radiative and non-radiative decay rates of the different modes,~\cite{Kante2021inpreparation} which go beyond our work.

As a final point, it is worth briefly mentioning some straightforward experiments that may serve to shed light on the possible interpretations of the experimental observations even in the absence of a direct measurement of the $Q$ factor of the different modes. In the IR spectral region of the experiment, magnetic effects are quite weak as signalled by the smallness of the topological gap. This implies that the magnetic field is crucial to induce the topological edge state, but has a minor effect on the bulk regions. As a result, according to our theory in the absence of any magnetic field no laser operation should be observed up to powers well above the topological laser threshold. Some evidence in this direction is found by comparing Figs.~3(b) and 3(c) of Ref.~\onlinecite{Bahari2017}. Further experimental insight could be obtained by keeping the magnetic field on and ramping up the pump intensity well above the topological laser threshold. According to our model, as discussed in Sec.~\ref{sec:broadband}, going up in gain strength $G$ should move the system from the topological lasing region indicated in blue into the ones of multimode bulk lasing indicated in red/yellow. In particular, we expect that the threshold for bulk lasing at high gain strengths should be almost insensitive to the applied magnetic field.

\section{Conclusions}
\label{sec:conclusions}

In this work, we have developed a general semiclassical theory of topological laser operation that is able to include the peculiar structure of the photonic modes of the underlying topological lattice and the frequency-dependence of a realistic gain material. As a specific example of application of our theory, we have investigated the lasing threshold in a configuration that displays a subtle interplay between the spatial overlap of the modes with the gain medium and the spectral position and width of the frequency gaps in the different regions. Based on our theory, we propose an interpretation of the recent experiments in Ref.~\onlinecite{Bahari2017}, where stable topolaser emission was observed in spite of the gain being distributed across the whole photonic crystal structure and not localized on the topological edge as in Ref.~\onlinecite{Bandres2018}.

A natural next step will be to include our theory of frequency-dependent gain into the Bogoliubov description of collective excitations around a topologically lasing state~\cite{Loirette-Pelous2021} so to characterize the stability of realistic models of topological laser operation in the different regimes of gain parameters. This will be of great interest as a new tool to tame all those instability mechanisms that may hinder a clean and efficient single-mode topological laser emission in practical semiconductor devices.

\section*{Supplementary Material}
\label{sec:supp}

See supplementary material for full derivations of the theoretical models, additional simulations at different gain widths and an extended discussion on the topological lasing features.

\acknowledgments

We warmly acknowledge continuous stimulating exchanges with Boubacar Kanté. We are grateful to Aurelian Loirette-Pelous and Ivan Amelio for continuous discussions on topolaser dynamics and we thank Maxim Gorlach for discussions on our preliminary results. We acknowledge financial support from the European Union H2020-FETFLAG-2018-2020 project "PhoQuS" (n.820392) and from the Provincia Autonoma di Trento. TO acknowledges support from JSPS KAKENHI Grant Number JP20H01845, JST PRESTO Grant Number JPMJPR19L2, and JST CREST Grant Number JPMJCR19T1. The numerical simulations were performed on SISSA's Ulysses cluster.

\section*{Data Availability}

The data that support the findings of this study are available from the corresponding author upon reasonable request.

\bibliographystyle{apsrev4-1}   
\typeout{}                      
\bibliography{references}

\includesupplementary

\end{document}


\title{Supplementary Information for\texorpdfstring{\\}{} ``Spatial and Spectral Mode-Selection Effects in Topological Lasers with Frequency-Dependent Gain''}

\author{Matteo Seclì}
\email[]{matteo.secli@sissa.it}
\affiliation{\mbox{International School for Advanced Studies (SISSA), Via Bonomea 265, I-34136 Trieste, Italy}}

\author{Tomoki Ozawa}
\affiliation{\mbox{Advanced Institute for Materials Research, Tohoku University, Sendai 980-8577, Japan}}

\author{Massimo Capone}
\affiliation{\mbox{International School for Advanced Studies (SISSA), Via Bonomea 265, I-34136 Trieste, Italy}}
\affiliation{CNR-IOM Democritos, Via Bonomea 265, I-34136 Trieste, Italy}

\author{Iacopo Carusotto}
\affiliation{\mbox{INO-CNR BEC Center and Dipartimento di Fisica, Università di Trento, I-38123 Povo, Italy}}

\date{April 15, 2021}

\maketitle

In this supplementary document we provide additional material in support of our conclusions. In particular, we give a detailed derivation of the semiclassical Bloch-Harper-Hofstadter model (Sec.~\ref{supp:sec:semiclassical_laser_theory}); we review general features of lasing with frequency-dependent gain (Sec.~\ref{supp:sec:mode_pull}); we perform a time-frequency analysis of the lasing process (Sec.~\ref{supp:sec:time_frequency_analysis}); we provide additional numerical results for different parameter choices (Sec.~\ref{supp:sec:extra_phase_diagrams}); we give analytical expressions for the transition lines between different regions in the phase diagram (Sec.~\ref{supp:sec:transition_lines}).

\section{Semiclassical Theory of Lasers}
\label{supp:sec:semiclassical_laser_theory}

\subsection{Model and Equations}
\label{supp:sec:semiclassical_laser_theory_equations}

Lamb's semiclassical theory of lasing is usually derived by coupling a semiclassical Hamilonian for the light-atom interaction, where the electric field is classical, with Maxwell's equations. The same theory can also be obtained under a mean-field treatment of the photon field, by starting from a fully quantum Hamiltonian, e.g.\@ the Jaynes-Cummings Hamiltonian; while in the former case one has to make the assumption of slowly-varying fields, in the quantum treatment we work under the dipole and the rotating wave approximations.

The starting point is an Hamiltonian description of $N$ identical two-level atoms (TLAs) inside a cavity; the choice of working with TLAs comes from the fact that they're the most simple --- yet effective --- model of an emitter. The multi-atom version of the Jaynes-Cummings Hamiltonian, also called Tavis-Cummings or Dicke model, reads as:~\cite{Seke1990}
\begin{align}
    H
    &= \hbar\omega_{\mathrm{cav}} a^{\dagger}a + \sum_{n=1}^N \Bigg\lbrace \frac{1}{2}\hbar\omega_{eg}\sigma^z_n
    + \hbar g \Big( \sigma^+_n a + a^{\dagger} \sigma^-_n \Big) \Bigg\rbrace \nonumber \\
    &\doteqdot H^{(r)} + \sum_{n=1}^NH^{(ar)}_n,
\end{align}
where $(r)$ stands for ``radiation'' and $(ar)$ for ``atom-radiation'' interaction. Here $\omega_{\mathrm{cav}}$ is the natural frequency of the cavity, $\omega_{eg} = \omega_e - \omega_g$ is the energy difference between the atomic levels, $g = g_{eg} = g_{ge} = -\frac{d_{eg}E_0}{\hbar}$ is the light-atom coupling, $d_{eg} = |\vec{d}_{eg}| = ez_{eg} = d_{eg}^*$, $\vec{d}_{eg} = \vec{d}_{ge}^*$ is the matrix element $\Braket{e|e\vec{r}|g}$ of the dipole moment ---  usually called the \emph{transition dipole moment}, $E_0 = \sqrt{\frac{\hbar\omega_{\mathrm{cav}}}{2\varepsilon_0V}}$ and $V$ is the volume of the cavity. In the Hamiltonian, $a^{\dagger}$ is an operator that creates a cavity photon and, for every atom $n$, $\sigma^+ = \ket{e}\bra{g}$ and $\sigma^- = (\sigma^+)^{\dagger} = \ket{g}\bra{e}$ are operators that describe transitions between the atomic levels, and $\sigma^z = \ket{e}\bra{e} - \ket{g}\bra{g}$. Additionally $\ket{e}\bra{e} + \ket{g}\bra{g} = I$, with $I$ the identity operator.

The central assumption we now make is for the total density matrix $\rho$ to be factorizable as a tensor product between a density matrix for the atoms and a density matrix for the photons, i.e.\@ $\rho = \rho_{N-\mathrm{at}}\otimes\rho_{\mathrm{ph}}$, where the atomic part can be obtained by tracing out the photonic part: $\rho_{N-\mathrm{at}} = \mathrm{Tr}_{\mathrm{ph}}\rho$. With this consideration one can similarly obtain a time-evolution equation for the atomic density matrix $\rho_{N-\mathrm{at}}$ by starting from the evolution of the full density matrix, $\frac{d\rho}{dt} = -\frac{i}{\hbar}\left[H, \rho\right]$, and tracing out the photonic part. The result is an equation very similar to the one for the full density matrix, but in which $H$ is replaced by a mean-field Hamiltonian:
\begin{align}
    \frac{d\rho_{N-\mathrm{at}}}{dt}
    &= -\frac{i}{\hbar}\left[H^{\mathrm{MF}}, \rho_{N-\mathrm{at}}\right] \nonumber \\
    &= -\frac{i}{\hbar}\sum_{n=1}^N\left[H^{\mathrm{MF},\,(ar)}_n, \rho_{N-\mathrm{at}}\right],
\end{align}
where in $H^{\mathrm{MF}}$ all the photon operators have been replaced by their expectation values:
\begin{equation}
    H^{\mathrm{MF}}
    = \hbar\omega_{\mathrm{cav}} \Braket{a^{\dagger}a}
    + \sum_{n=1}^N \Bigg\lbrace \frac{1}{2}\hbar\omega_{eg}\sigma^z_n
    + \hbar g \Big( \sigma^+_n \Braket{a} + \Braket{a^{\dagger}} \sigma^-_n \Big) \Bigg\rbrace
\end{equation}

If we further assume that the atomic density matrix can be factorized as a tensor product of the density matrices of all the atoms, i.e.\@
\begin{equation}
    \rho_{N-\mathrm{at}} = \rho_{\mathrm{at}}^{(1)} \otimes \ldots \otimes \rho_{\mathrm{at}}^{(N)}
\end{equation}
where $\rho_{\mathrm{at}}^{(n)} = \sum_{\{\alpha,\,\beta\} = \{e,\,g\}} \rho_{\alpha\beta}^{(n)}\ket{\alpha}\bra{\beta}$ is the density matrix of the $n$-th atom, then the commutator above simplifies as
\begin{equation}
    \left[H^{\mathrm{MF},(ar)}_n, \rho_{N-\mathrm{at}}\right]
    = \left[H^{\mathrm{MF},(ar)}_n, \rho_{\mathrm{at}}^{(n)}\right].
\end{equation}

If we trace away all the other atoms except for the $n$-th one, we get
\begin{equation}
    \frac{d\rho_{\mathrm{at}}^{(n)}}{dt} = 
    -\frac{i}{\hbar}\left[H^{\mathrm{MF},(ar)}_n, \rho_{\mathrm{at}}^{(n)}\right]
    \label{supp:eq:drhoat_simplified}
\end{equation}
which is exactly the same time-evolution equation one would get from a single-atom Jaynes-Cummings Hamiltonian. This means that all the atoms are equivalent and have exactly the same evolution, so we can drop the $(n)$ superscript. 

Finally, we can provide every atom with a decay + pumping mechanism at the master equation level, by adding Lindblad terms of the following form:
\begin{equation}
    \frac{d\rho_{\mathrm{at}}}{dt} = 
    \ldots + \sum_{s=e,g}\gamma_s\left( L_s\rho_{\mathrm{at}}L_s^{\dagger} -\frac{1}{2}\left\lbrace L_s^{\dagger}L_s, \rho_{\mathrm{at}}\right\rbrace \right),
\end{equation}
where the dots $\ldots$ indicate the rhs of \eqref{supp:eq:drhoat_simplified}, $L_e^{\dagger} = \Ket{e}\Bra{g} = \sigma^+$ (the population of $\Ket{e}$ is increased) and $L_e = \Ket{g}\Bra{e} = \sigma^-$ (the population of $\Ket{e}$ is decreased), and conversely $L_g^{\dagger} = \Ket{g}\Bra{e} = \sigma^-$ and $L_g = \Ket{e}\Bra{g} = \sigma^+$.

With these ingredients we can write down the equations for the matrix elements of the atomic density matrix of a generic atom in the cavity:
\begin{equation}
    \begin{cases}
    \displaystyle \dot{\rho}_{ee} = \gamma_g\rho_{gg} - \gamma_e\rho_{ee} +i\frac{d_{eg}E_0}{\hbar}\Big( \braket{a}\rho_{ge} - \braket{a}^*\rho_{eg} \Big) \\
    \displaystyle \dot{\rho}_{gg} = \gamma_e\rho_{ee} - \gamma_g\rho_{gg} -i\frac{d_{eg}E_0}{\hbar}\Big( \braket{a}\rho_{ge} - \braket{a}^*\rho_{eg} \Big) \\
    \displaystyle \dot{\rho}_{eg} = -i(\omega_{eg} - i\gamma)\rho_{eg} - i\frac{d_{eg}E_0}{\hbar}\braket{a}\Big( \rho_{ee} - \rho_{gg} \Big)
    \end{cases}
    \label{supp:eq:TLA_OBE_phonon}
\end{equation}
where $\gamma = \gamma_{eg} + \gamma_{\mathrm{ph}}$, $\gamma_{eg} = \frac{\gamma_e+\gamma_g}{2}$ and $\gamma_{\mathrm{ph}}$, in the more general treatment (see Ref.~\onlinecite{Scully1997}), comes from dephasing processes; for the sake of our discussion though, $\gamma_{\mathrm{ph}} = 0$. The fact that the evolution of the population of one of the levels is the negative of the evolution of the other one is due to the fact that the Lindblad losses preserve the trace of the density matrix (which is $1$), and indeed from $\rho_{gg} + \rho_{ee} = 1$ we get $\dot{\rho}_{gg} = -\dot{\rho}_{ee}$.

We still miss an equation for the cavity field $a$, or better, for its expectation value. The Heisenberg equation for the field operator is
\begin{equation}
    \frac{da}{dt}
    = \frac{i}{\hbar}\left[H, a\right]
    = -i\omega_{\mathrm{cav}} a - ig\sum_{n=1}^N\sigma^-_n.
    \label{supp:eq:cavity_Heisenberg}
\end{equation}
By noticing that $\sigma^-$ is an atom-only operator, so that
\begin{equation}
    \braket{\sigma^-} = \braket{\sigma^-}_{\mathrm{at}} = \mathrm{Tr}_\mathrm{at}\Big(\rho_{\mathrm{at}}\sigma^-\Big) = \rho_{eg},
\end{equation}
we can then take the expectation value of \eqref{supp:eq:cavity_Heisenberg} and use the fact the the atoms are all equivalent to write 
\begin{align}
    \frac{d\Braket{a}}{dt}
    &= -i\omega_{\mathrm{cav}} \Braket{a} - ig\sum_{n=1}^N\rho_{eg}^{(n)} \nonumber \\
    &= -i\omega_{\mathrm{cav}} \Braket{a} - igN\rho_{eg}.
    \label{supp:eq:Ma_MF}
\end{align}

Finally, by redefining $\braket{a} \to a$ for simplicity and by adding a phenomenological cavity loss coefficient $\Gamma$, we have a set of coupled semi-classical equations for the elements of the atomic density matrix and the expectation value of the photon field:
\begin{equation}
    \begin{cases}
    \displaystyle \dot{\rho}_{ee} = \gamma_g\rho_{gg} - \gamma_e\rho_{ee} +i\frac{d_{eg}E_0}{\hbar}\Big( a\rho_{ge} - a^*\rho_{eg} \Big) \\
    \displaystyle \dot{\rho}_{gg} = \gamma_e\rho_{ee} - \gamma_g\rho_{gg} -i\frac{d_{eg}E_0}{\hbar}\Big( a\rho_{ge} - a^*\rho_{eg} \Big) \\
    \displaystyle \dot{\rho}_{eg} = -i(\omega_{eg} - i\gamma)\rho_{eg} -i\frac{d_{eg}E_0}{\hbar}a\Big( \rho_{ee} - \rho_{gg} \Big)\\
    \displaystyle \dot{a} = -i\omega_{\mathrm{cav}} a -\Gamma a +i\frac{d_{eg}E_0}{\hbar}N\rho_{eg}
    \end{cases}
    \label{supp:eq:MTLA_equations}
\end{equation}

Interestingly the number of atoms in the cavity, $N$, appears only in the equation for the average photon field, in front of the off-diagonal atomic matrix element. Indeed, if you compare with the semiclassical theory,~\cite{Sargent1974,Scully1997} you find that the time-evolution for the macroscopic electric field is
\begin{equation}
    \dot{E}^{+}
    = -i\omega_{\mathrm{cav}}E^{+} + i\frac{\omega_{\mathrm{cav}}}{2\varepsilon_0}P^{+},
\end{equation}
where $E^+$ is the positive-frequency component of the electric field and $P^+$ is the positive-frequency component of the (macroscopic) polarization. Each atom here has a tiny dipole moment that can be shown to be $\braket{e\vec{r}} = d_{eg}(\rho_{eg} + \rho_{ge}) = d_{eg}(\rho_{eg} + \rho_{eg}^*)$, so one can assume that the macroscopic polarization vector is just the sum of these tiny dipole moments:
\begin{equation}
    P = N\braket{e\vec{r}} = Nd_{eg} \Big( \rho_{eg} + \text{c.c.} \Big).
\end{equation}
When you then identify $P^+ \equiv Nd_{eg} \rho_{eg}$, you get 
\begin{equation}
    \dot{E}^{+}
    = -i\omega_{\mathrm{cav}}E^{+} - ig'N\rho_{eg}
\end{equation}
where $g'=-\frac{\omega_{\mathrm{cav}}d_{eg}}{2\varepsilon_0}$, which has the same form of \eqref{supp:eq:Ma_MF}.

\subsection{Stationary State}
\label{supp:sec:semiclassical_laser_theory_ss}

We can get additional insights and build up some intuition by solving the equations \eqref{supp:eq:MTLA_equations} for the \emph{stationary state}.

In order to do that, let us first perform a change of variable. We define the overall oscillation frequency of the field as $\omega_L$ --- which is not \textit{a priori} guaranteed to be equal to $\omega_{\mathrm{cav}}$, so that it is convenient to define a new variable $\tilde{a}$ as
\begin{equation}
    a(t) \doteqdot \tilde{a}(t)e^{-i\omega_Lt},
    \label{supp:eq:varchange_wL}
\end{equation}
where now $\tilde{a}(t)$ is a \emph{slowly-varying} quantity, i.e.\@ $|\dot{\tilde{a}}/\tilde{a}| \ll \omega_L$.  We also define $\tilde{\rho}_{eg} = \rho_{eg}e^{+i\omega_Lt}$ and $\tilde{\rho}_{\alpha\alpha} \equiv \rho_{\alpha\alpha}$ ($\alpha=e,g$). The steady-state coherence in \eqref{supp:eq:MTLA_equations} can then be expressed as
\begin{equation}
	\tilde{\rho}_{eg} = g \tilde{a} \frac{\Big[(\omega_{eg}-\omega_L) + i\gamma\Big]}{\Big[(\omega_{eg}-\omega_L)^2 + \gamma^2\Big]} \Big( \tilde{\rho}_{ee} - \tilde{\rho}_{gg} \Big), \\
	\label{supp:eq:MTLA_ss_coherence}
\end{equation}
while the steady-state population difference takes the form
\begin{equation}
	\Big( \tilde{\rho}_{ee} - \tilde{\rho}_{gg} \Big) = \frac{\Big( \tilde{\rho}_{ee}^{(0)} - \tilde{\rho}_{gg}^{(0)} \Big)}{1 + R/\gamma_{ul}}\,,
	\qquad\qquad
	R = 2g^2|\tilde{a}|^2 \frac{\gamma}{(\omega_{eg}-\omega_L)^2 + \gamma^2}\,,
	\label{supp:eq:MTLA_ss_pop_diff}
\end{equation}
where $\tilde{\rho}_{ee}^{(0)} = \frac{\gamma_g}{\gamma_g+\gamma_e}$ and $\tilde{\rho}_{gg}^{(0)} = \frac{\gamma_e}{\gamma_g+\gamma_e}$ are the steady-state populations at zero field.

By combining \eqref{supp:eq:MTLA_ss_coherence} and \eqref{supp:eq:MTLA_ss_pop_diff}, the equation for the field can be written in the following form:
\begin{equation}
    \dot{\tilde{a}} 
    = i\Omega\tilde{a} - \Gamma\tilde{a} + \frac{P}{1+\beta|\tilde{a}|^2}\tilde{a}
\end{equation}
where
\begin{align}
    \Omega 
    = \Omega(\{\omega\},\{\gamma\},|\tilde{a}|^2)
    = (\omega_L-\omega_{\mathrm{cav}}) - g^2N \frac{(\omega_{eg}-\omega_L)}{\Big[(\omega_{eg}-\omega_L)^2 + \gamma^2\Big]}\frac{\Big( \tilde{\rho}_{ee}^{(0)} - \tilde{\rho}_{gg}^{(0)} \Big)}{1+\beta|\tilde{a}|^2},
\end{align}
\begin{align}
     P
     = P(\omega_{eg}-\omega_L,\{\gamma\})
     = g^2N\frac{\gamma}{\Big[(\omega_{eg}-\omega_L)^2 + \gamma^2\Big]}\Big( \tilde{\rho}_{ee}^{(0)} - \tilde{\rho}_{gg}^{(0)} \Big)
\end{align}
and
\begin{equation}
    \beta
    = \beta(\omega_{eg}-\omega_L,\gamma)
    = 2g^2 \frac{\gamma/\gamma_{eg}}{(\omega_{eg}-\omega_L)^2 + \gamma^2}.
    \label{supp:eq:sat_beta}
\end{equation}
Here $\{\omega\}$ indicates the dependence on the frequencies ($\omega_{\mathrm{cav}}$, $\omega_{eg}$ and $\omega_L$) and $\{\gamma\}$ indicates the dependence on the microscopic loss coefficients ($\gamma_e,\gamma_g,\gamma_{\mathrm{ph}}$). 

As a key point, the $P$ coefficient which, for a positive zero-field population difference $\Big( \tilde{\rho}_{ee}^{(0)} - \tilde{\rho}_{gg}^{(0)} \Big)=(\gamma_g-\gamma_e)/(\gamma_g+\gamma_e)>0$, acts as a \emph{gain}, is here \emph{frequency-dependent}; you get most out of your cavity system when the lasing frequency $\omega_L$ is in resonance with the transition frequency between the two atomic levels $\omega_{eg}$. If instead your lasing frequency gets farther apart from the  transition frequency, then the gain decays by following a Lorentzian behavior --- with the linewidth of the Lorentzian controlled by $\gamma$. Additionally, the lasing frequency experiences an intensity-dependent frequency shift that decays as the square of the detuning for large detunings but is exactly zero at resonance; this shift is responsible for the so-called \emph{mode pulling}.

\subsection{Active Harper-Hofstadter Lattice}
\label{supp:sec:harper_hofstadter}

Multiple active cavities like the one studied in Sec.~\ref{supp:sec:semiclassical_laser_theory} can be further arranged into a lattice with a suitably engineered hopping Hamiltonian. In our case, we are interested in a Harper-Hofstadter model with a hopping term of the form:
\begin{equation}
    H_{\mathrm{hop}} = -J\sum_{m,n} \Big\lbrace a_{m,n}^{\dagger}a_{m+1,n} + e^{-i 2\pi\vartheta m}a_{m,n}^{\dagger}a_{m,n+1} + \text{h.c.} \Big\rbrace,
    \label{supp:eq:HH_tight_binding_app}
\end{equation}
where $m=1,\ldots,N_x$ and $n=1,\ldots,N_y$ are the site indices in the $x$ and $y$ direction of the lattice plane. The introduction of a hopping Hamiltonian amounts to an additional term in the evolution of the fields (third equation in \eqref{supp:eq:MTLA_equations}), i.e.\@ the photon field at each site, before taking the expectation value, now evolves as
\begin{equation}
    \dot{a}_{m,n} 
    = \ldots + \frac{i}{\hbar}\left[H_{\mathrm{hop}}, a_{m,n}\right],
    \label{supp:eq:Heisenberg_tight_binding}
\end{equation}
where the dots $\ldots$ indicate the rhs of \eqref{supp:eq:cavity_Heisenberg}.

It is also useful to perform a change of variable so to measure all frequencies with respect to the cavity frequency $\omega_{\mathrm{cav}}$. Note that this is different from the transformation done in \eqref{supp:eq:varchange_wL}; while the latter is useful for a theoretical discussion, is of little help in the numerical simulations since the lasing frequency $\omega_L$ is unknown. In practice, we rewrite $a(t)$ in the following way:
\begin{align}
    a(t) 
    = a_0(t)e^{-i\omega_Lt}
    = \underbrace{a_0(t)e^{-i(\omega_L-\omega_{\mathrm{cav}})t}}_{\tilde{\alpha}(t)}e^{-i\omega_{\mathrm{cav}}t}
    = \tilde{\alpha}(t)e^{-i\omega_{\mathrm{cav}}t}
    \label{supp:eq:atilde_definition}
\end{align}
where now $\tilde{\alpha}(t)$ is \emph{not} a slowly varying quantity, since it oscillates at the detuning between the field and the cavity. We further define $\tilde{\rho}_{eg} = \rho_{eg}e^{i\omega_{\mathrm{cav}}t}$, $\tilde{\rho}_{nn} \equiv \rho_{nn}$ ($n=e,g$), $\tilde{\omega}_{eg} = \omega_{eg}-\omega_{\mathrm{cav}}$ and $\tilde{\omega}_L = \omega_L-\omega_{\mathrm{cav}}$, with the consideration that equation \eqref{supp:eq:Heisenberg_tight_binding} holds for the transformed field $\tilde{\alpha}$ as well.

For numerical purposes, it is also useful to perform a rescaling to dimensionless quantities. This is done by performing the re-definitions $\frac{\hbar\gamma_{\{e,g,eg,\mathrm{ph}\}}}{J} \to \gamma_{\{e,g,eg,\mathrm{ph}\}}$, $\frac{\hbar\tilde{\omega}_{eg}}{J} \to \tilde{\omega}_{eg}$, $\frac{Jt}{\hbar} \to t$, $\frac{\hbar\Gamma}{J} \to \Gamma$, $-\frac{\hbar g}{J}\tilde{\alpha} \to \tilde{\alpha}$ (the extra ``$-$'' compensates the fact that $g$ is negative), and by defining the new quantity
\begin{equation}
G \doteqdot \left(\frac{\hbar g}{J}\right)^2N = \left(\frac{d_{eg} E_0}{J}\right)^2N = \frac{d_{eg}^2\hbar\omega_{\mathrm{cav}}}{2J^2\varepsilon_0}\frac{N}{V}.
\end{equation}
This parameter, which is proportional to the volumetric density $\frac{N}{V}$ of TLAs and to the natural frequency of the cavity, acts as a \emph{gain parameter}.

With these definitions, and by dropping all the remaining ``\char`\~'' for simplicity, the final equations for a lattice of Harper-Hofstadter cavities are:
\begin{equation}
    \begin{cases}
    \displaystyle \dot{\rho}_{ee}^{m,n} = \gamma_g - 2\gamma_{eg}\rho_{ee}^{m,n} + i\Big( \alpha_{m,n}\rho_{ge}^{m,n} - \alpha^{*}_{m,n}\rho_{eg}^{m,n} \Big) \\
    \displaystyle \dot{\rho}_{gg}^{m,n} = \gamma_e - 2\gamma_{eg}\rho_{gg}^{m,n} - i\Big( \alpha_{m,n}\rho_{ge}^{m,n} - \alpha^{*}_{m,n}\rho_{eg}^{m,n} \Big) \\
    \displaystyle \dot{\rho}_{eg}^{m,n} = -i\Big(\omega_{eg} - i\gamma\Big)\rho_{eg}^{m,n} - i\alpha_{m,n}\Big( \rho_{ee}^{m,n} - \rho_{gg}^{m,n} \Big) \\
    \begin{aligned}[c]%
        \displaystyle \dot{\alpha}_{m,n} 
        = -\Gamma\alpha_{m,n} + iG\rho_{eg}^{m,n} + i\Big( \alpha_{m+1,n} + \alpha_{m-1,n} \Big)
        + i\Big( e^{-i 2\pi\vartheta m}\alpha_{m,n+1} + e^{+i 2\pi\vartheta m}\alpha_{m,n-1} \Big)
        \end{aligned}
    \end{cases}
    \label{supp:eq:topolaser_TLA_equations}
\end{equation}

In all the simulations, we've set the on-site losses to a reasonable value $\Gamma = 0.1$.

\subsection{Lasing with Frequency-Dependent Gain}
\label{supp:sec:harper_hofstadter_lasing}

We now seek the lasing condition of a single lattice cavity, bearing in mind that as we have shown in Ref.~\onlinecite{Secli2019} the lasing condition for a lattice of identical cavities has to be greater or equal than the threshold of a single cavity.

In actual devices the amplification is controlled by $\gamma_g$, the rate at which the atoms are excited from their ground state $\Ket{g}$ to a higher-energy state $\Ket{e}$. However, this has also the additional side effect of possibly modifying both the height and the linewidth of the Lorentzian profile of the gain (see Sec.~\ref{supp:sec:semiclassical_laser_theory_ss}), depending on the relative value of $\gamma_{\mathrm{ph}}$, as well as contributing to frequency-dependent shifts of the lasing frequency itself and of the saturation factor of the amplification term. In order to skim these side effects off the model and obtain a simpler, yet still meaningful description, we'll instead use $G$ as the primary amplification parameter, and we'll write the lasing condition for $G$ itself. Increasing $G$ instead of $\rho_{eg}$ (through $\gamma_g$) has the same effect on the amplification term in equation for $\tilde{a}$, but helps avoiding the effects described above and, if the range of probed $G$ values is not too large, it still gives a faithful description.

The lasing condition in terms of $G$ is easily obtained from the steady-state analysis done in Sec.~\ref{supp:sec:semiclassical_laser_theory_ss}, which shows we must have $\frac{P}{1+\beta|\tilde{a}|^2} > \Gamma$. By expanding this expression, we have the threshold condition
\begin{equation}
    G > \frac{2\Gamma\gamma_{eg}/\gamma}{\gamma_g-\gamma_e}\Big[\gamma^2 + \Delta_{\mathrm{det}}^2\Big],
    \label{supp:eq:LT_condition_NORES}
\end{equation}
where $\Delta_{\mathrm{det}} = \omega_{eg}-\omega_L$ measures the detuning of the lasing frequency $\omega_L$ from the atomic transition frequency $\omega_{eg}$. This relation shows that the threshold for a given mode has a quadratic dependence on its frequency, and that the mode that lases first --- i.e.\@ for which the threshold is as small as possible --- is the one at resonance, $\omega_L = \omega_{eg}$. In this resonant case, the threshold condition becomes
\begin{equation}
    G > \frac{2\gamma_{eg}\gamma\Gamma}{\gamma_g-\gamma_e} \doteqdot G_{\mathrm{res},0},
    \label{supp:eq:LT_condition_RES}
\end{equation}
where we have defined $G_{\mathrm{res},0}$ as the \emph{single-cavity resonant lasing threshold}.

In the numerical simulations, for the atomic parameters we have set $\gamma_{\mathrm{ph}} = 0$ and we have taken $\gamma_g = \frac{4}{3}\gamma$ and $\gamma_g = \frac{2}{3}\gamma$, with $2\gamma = \gamma_g + \gamma_e$ being the FWHM of the Lorentzian profile of the gain. With this choice of parameters, we then have $G_{\mathrm{res},0} = 3\gamma\Gamma$. The remaining parameters are changed throughout the text.

\section{Mode Pulling and Frequency Discretization}
\label{supp:sec:mode_pull}

\begin{figure}
    \centering
    \mbox{
        \hspace*{-21pt}\includegraphics[scale=0.5,valign=b]{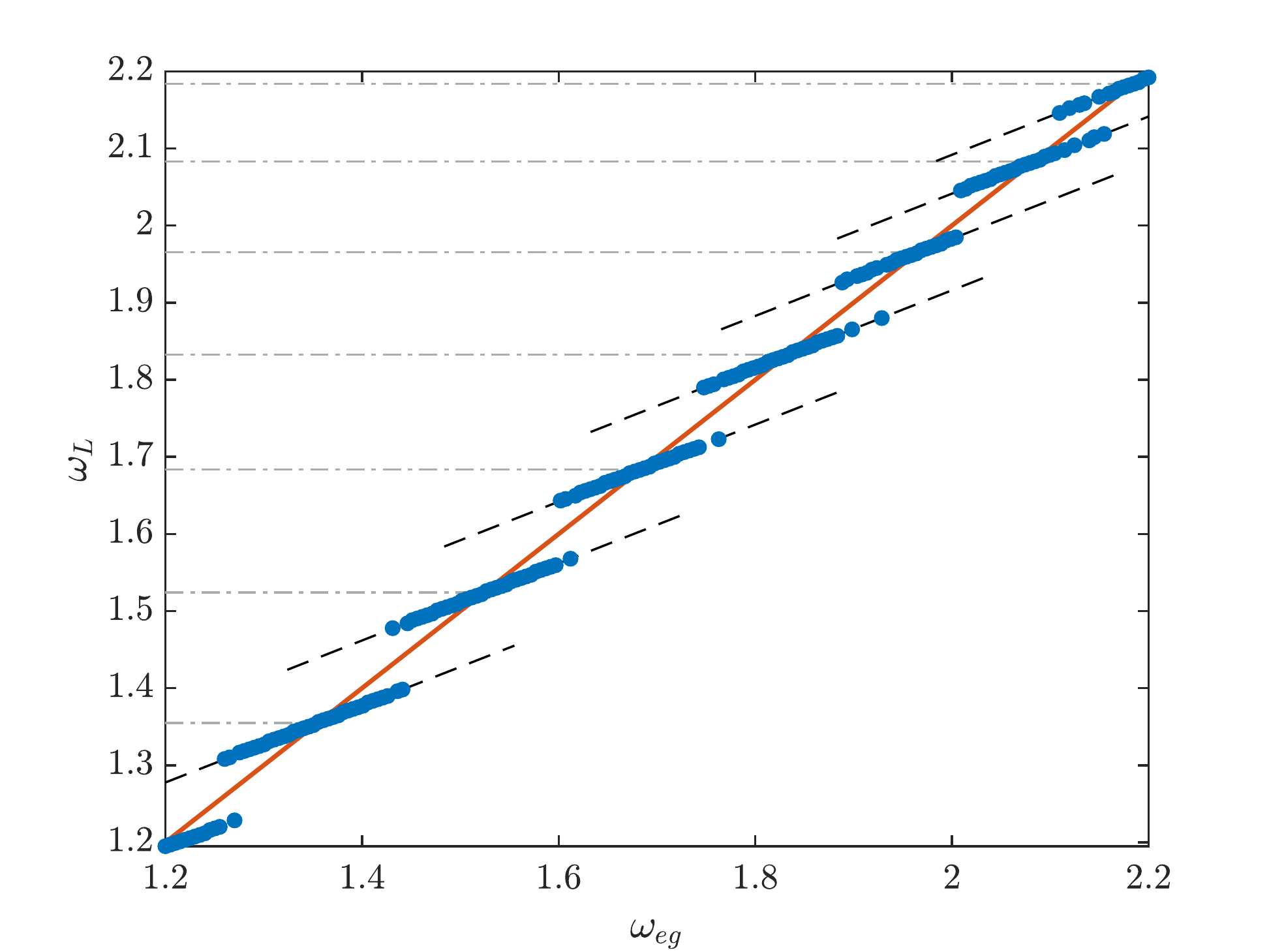}
        \hspace*{-12pt}\includegraphics[scale=0.5,valign=b]{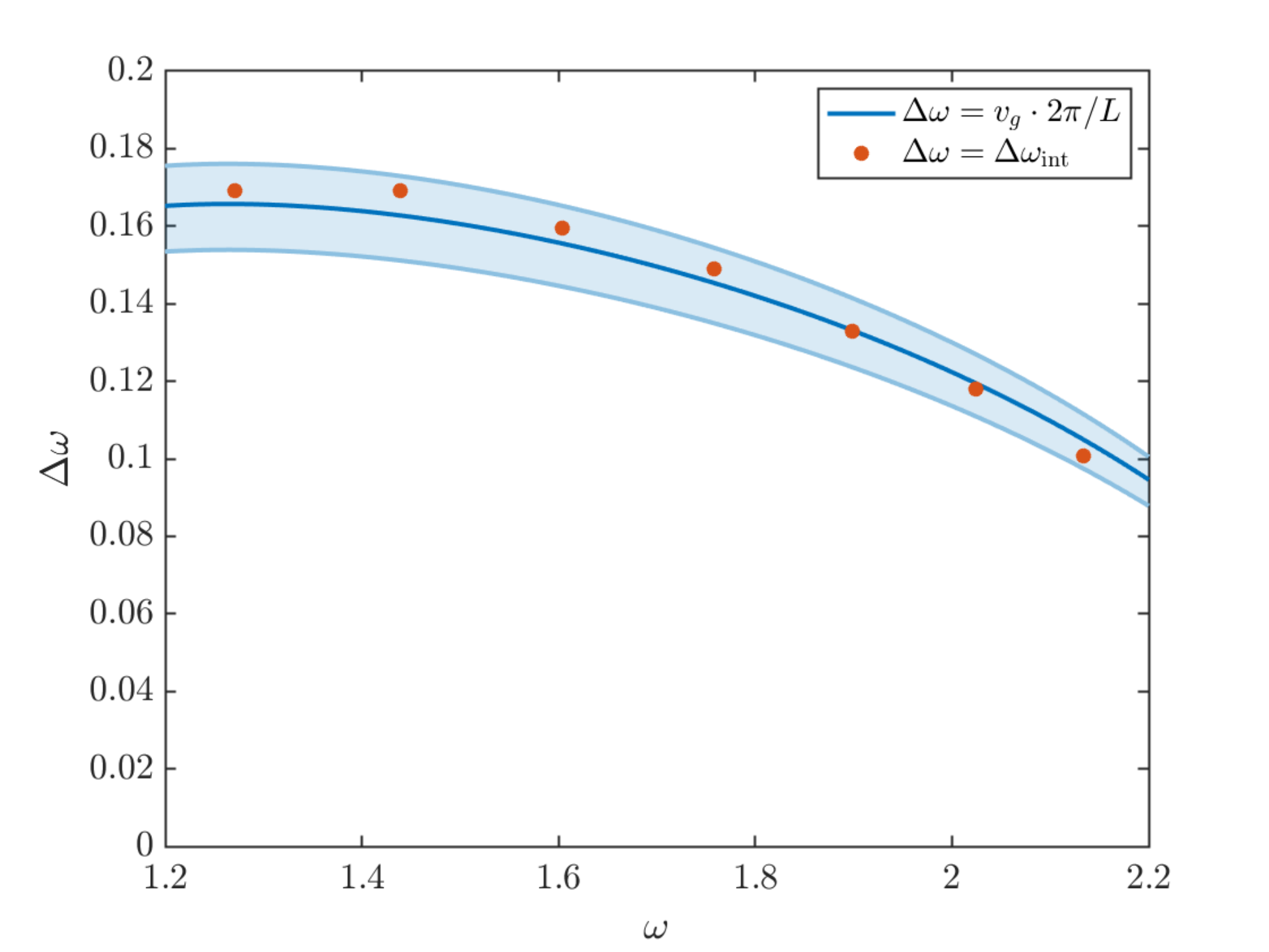}
    }
    \caption{Numerical simulations of topological laser operation in a $15 \times 15$ Harper-Hofstadter lattice and $\Gamma = \gamma = 0.1$.
    In the left panel, the blue dots represent the measured lasing frequencies obtained from a power spectral density analysis for different values of $\omega_{eg}$. Each slant step is fitted with a black, dashed line with equation $\omega_L = \frac{1}{2}\omega_{eg} + b$. The light grey dash-dot lines mark the lasing frequency values for which $\omega_L = \omega_{eg} = \omega_0$ (see also \eqref{supp:eq:MP_single_cavity}), while the red line represents the approximated curve $\omega_L = \omega_{eg}$.
    In the right panel, we compare the mode spacing $\Delta\omega$ calculated from the difference between the intersections $\omega_{\mathrm{int}}$ of adjacent steps in the left panel (red dots) with the one obtained from the group velocity (blue line) as discussed in the text.}
    \label{supp:fig:narrowband_mode_pulling}
\end{figure}

It is a general feature of laser physics that a narrowband gain causes an additional \emph{mode pulling} effect. For a single cavity this means that the lasing frequency \emph{does not} correspond to the bare cavity frequency, as it is pulled towards the center of the Lorentzian profile of the gain,~\cite{Scully1997} i.e.\@ towards $\omega_{eg}$:
\begin{equation}
    \omega_L = \frac{\omega_{\mathrm{cav}} + \mathcal{S}\omega_{eg}}{1+\mathcal{S}},
    \label{supp:eq:MP_single_cavity}
\end{equation}
where $\mathcal{S} = \frac{\Gamma}{\gamma}$ is called the \emph{stabilization factor}. For equal  $\Gamma = \gamma$, one has $\mathcal{S}=1$ and the mode pulling effect becomes a simple average.

In the case of a lattice of cavities, we have to replace $\omega_{\mathrm{cav}}$ with a frequency $\omega_{0}$ that corresponds to the specific mode of the underlying lattice that the system is selecting for lasing. This effect has to be considered for a reliable calculation of frequency-dependent lasing thresholds and to correctly relate the lasing frequencies that we measure through a power spectral density analysis to the underlying lattice eigenmodes.

Let us explore the impact of this effect on topolaser operation for $\omega_{eg}$ located in the topological bandgap. In this case, one may naively expect that mode pulling effects are irrelevant. The system is in fact expected to lase in the topological edge mode resonant with the gain for which the threshold is lowest. Since it is resonant, this mode does not experience any mode pulling --- setting $\omega_{0} = \omega_{eg}$ yields in fact $\omega_L = \frac{\omega_{0} + \mathcal{S}\omega_{eg}}{1+\mathcal{S}} = \omega_{eg}$. In contrast to this expectation, the numerical experiment shown in Fig.~\ref{supp:fig:narrowband_mode_pulling} shows that we actually have a richer physics due to the intrinsic discreteness of the modes. 

In the left panel, reproduced here from Fig.~2(d) of the main text, we have simulated a lattice with narrowband gain for multiple values of $\omega_{eg}$ and we have measured the steady-state frequency at which the system was emitting via a power spectral density analysis. While the measured frequencies obey an \emph{overall} approximate relation $\omega_L \sim \omega_{eg}$, that holds exactly only when $\omega_{eg}$ coincides with a lattice mode frequency $\omega_0$ (light grey dash-dot lines), the reality is that $\omega_L$ is bound to assume only \emph{discretized} values due to the periodic boundary conditions, as discussed in Ref.~\onlinecite{Secli2019}. In a broadband gain regime where mode pulling effects are negligible ($\mathcal{S}\ll 1$), this would result in $\omega_L$ following a stair-like behavior as a function of $\omega_{eg}$ with uniform spacing $\Delta\omega = v_g\cdot\frac{2\pi}{L}$ determined by the length $L$ of the topological edge. Within each step, the emission frequency is locked to the one of the mode that is closest to resonance and for which gain is strongest. For a narrowband gain, the behavior is instead that of a \emph{slant} stair and the slope of each of these slant steps is exactly $\frac{\mathcal{S}}{1+\mathcal{S}}$ (dashed lines). This results from the fact that $\omega_{0}$ \emph{cannot} continuously follow $\omega_{eg}$, but it has to do so in steps, so it is actually \emph{constant} in a certain frequency interval; then $\omega_L = \frac{\omega_{0} + \mathcal{S}\omega_{eg}}{1+\mathcal{S}}$ yields a behavior of type $y = a \cdot x + b$ with $a \equiv \frac{\mathcal{S}}{1+\mathcal{S}}$ and $b \equiv \frac{\omega_{0}}{1+\mathcal{S}}$. In the particular case $\mathcal{S} = 1$, we get $a = 1/2$ and $b = \omega_{0}/2$.

This also gives us the chance to quantitatively verify the relation between the spacing of the lasing frequencies and the group velocity of the topological edge mode in the presence of mode pulling (Fig.~\ref{supp:fig:narrowband_mode_pulling}, right panel). We can calculate the spacing between successive modes by just calculating the intersection frequencies $\omega_{\mathrm{int}}$ between the fits of type $y = \frac{1}{2}\cdot x +b $ (dashed lines) and the line $y = x$ (red line) and then taking the difference $\Delta \omega_{\mathrm{inc}}$ between successive frequencies. We assign each value $\Delta \omega_{\mathrm{int}}$ to the frequency at the center of its corresponding frequency interval; this first order approximation becomes exact for $L \to \infty$. From our periodic boundary condition argument, we should also have that the spacing between successive modes is $\Delta\omega \sim v_g \cdot\frac{2\pi}{L}$; as it's not clear how to treat the corner sites in the calculation of $L$, we show values ranging from $L = 4 \times 14$ to $L = 4 \times 16$ as a blue shaded area in the plot, with the solid blue line marking the value $L = 4 \times 15$.

The data shows a good agreement with respect to our predictions, highlighting the importance of including mode-pulling effects as well as the presence of mode discretization even in the narrowband case.

\section{Time-Frequency Analysis of the Lasing Process}
\label{supp:sec:time_frequency_analysis}

\begin{figure}[htbp]
    \centering
    \mbox{
        \hspace*{-24pt}\includegraphics[scale=0.5,valign=b]{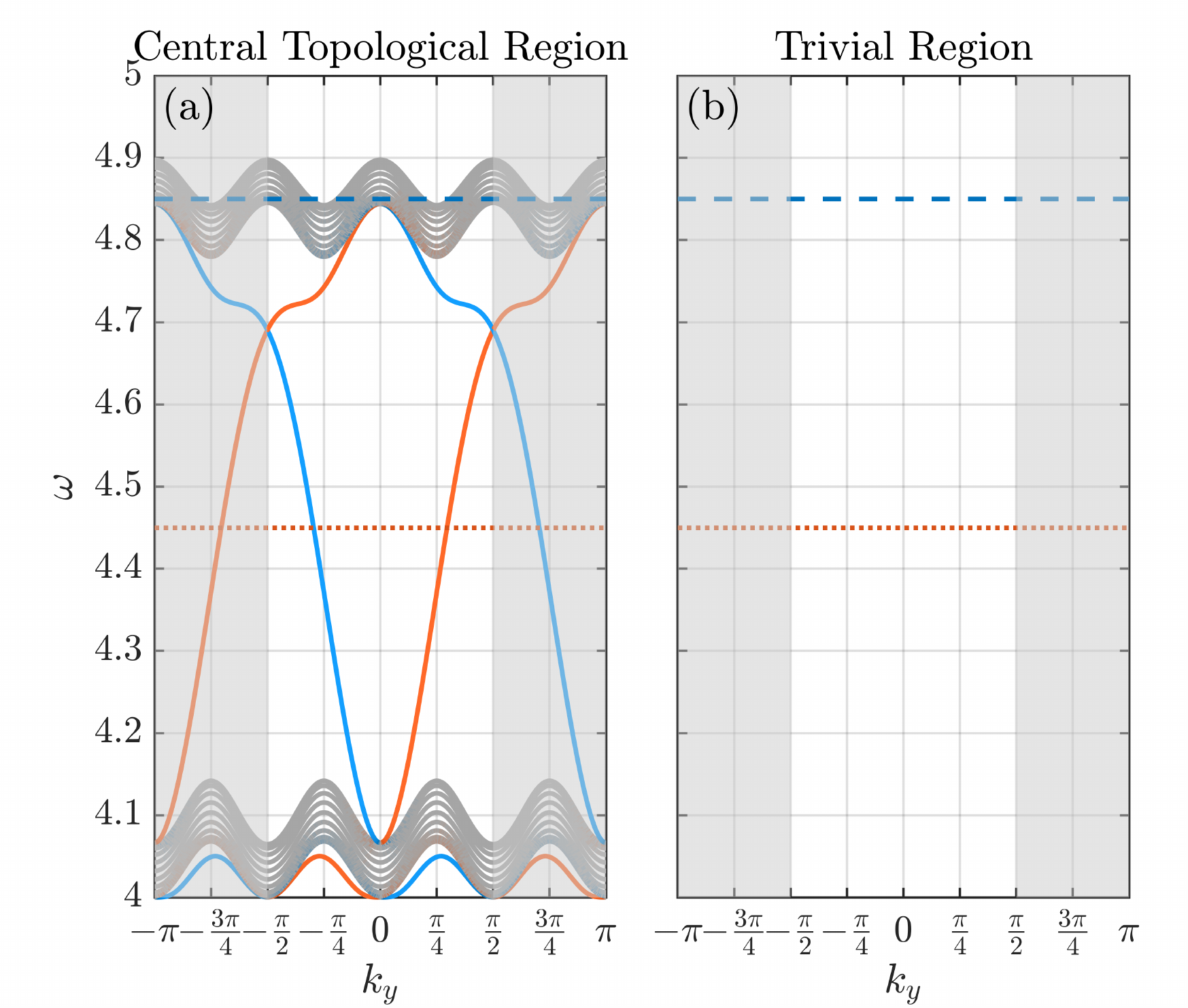}\hspace*{28.5pt}
        \hspace*{-6pt}\includegraphics[scale=0.5,valign=b]{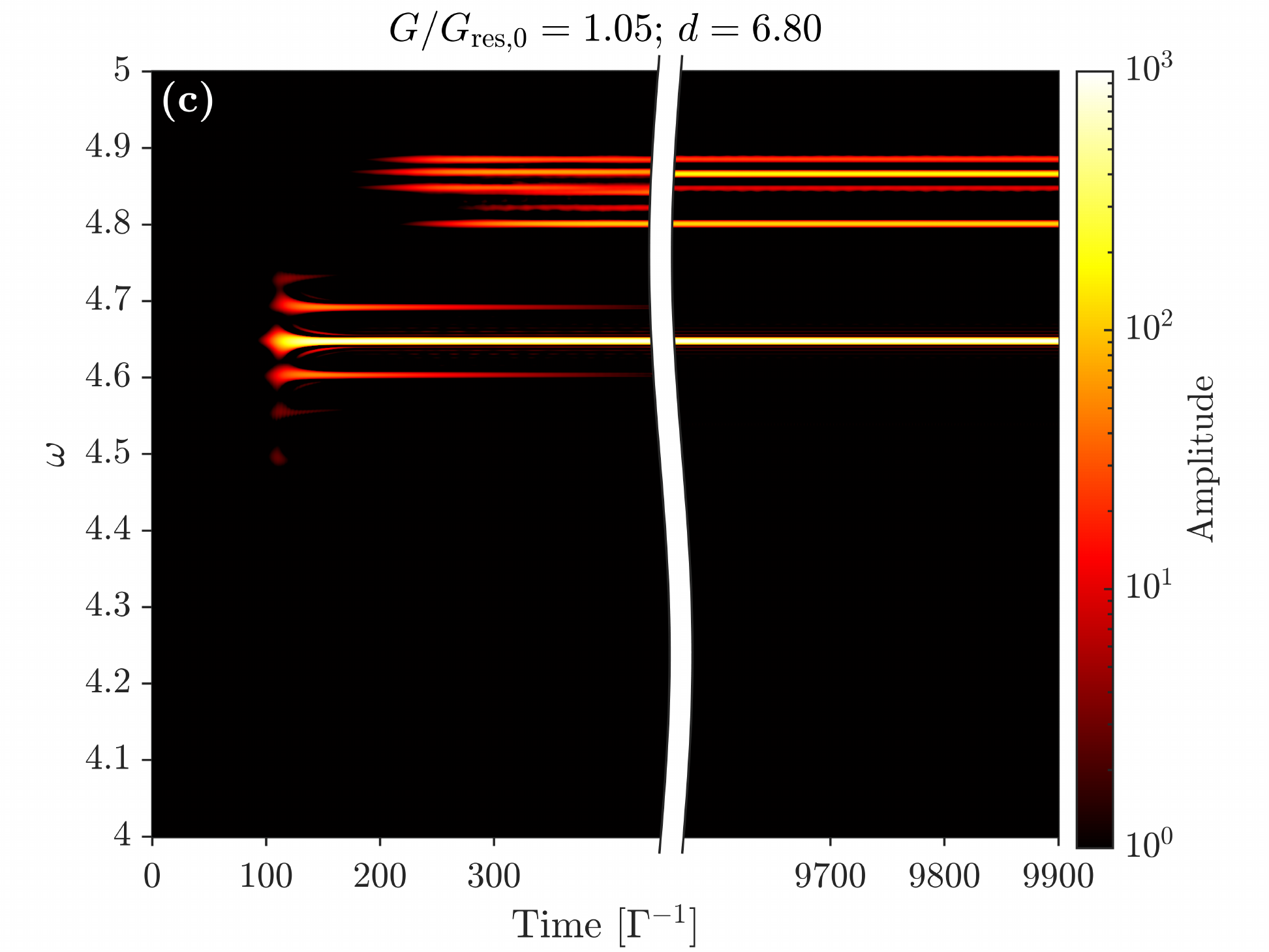}
    }
    \null\vspace*{6pt}
    \mbox{
        \hspace*{-24pt}\includegraphics[scale=0.5,valign=b]{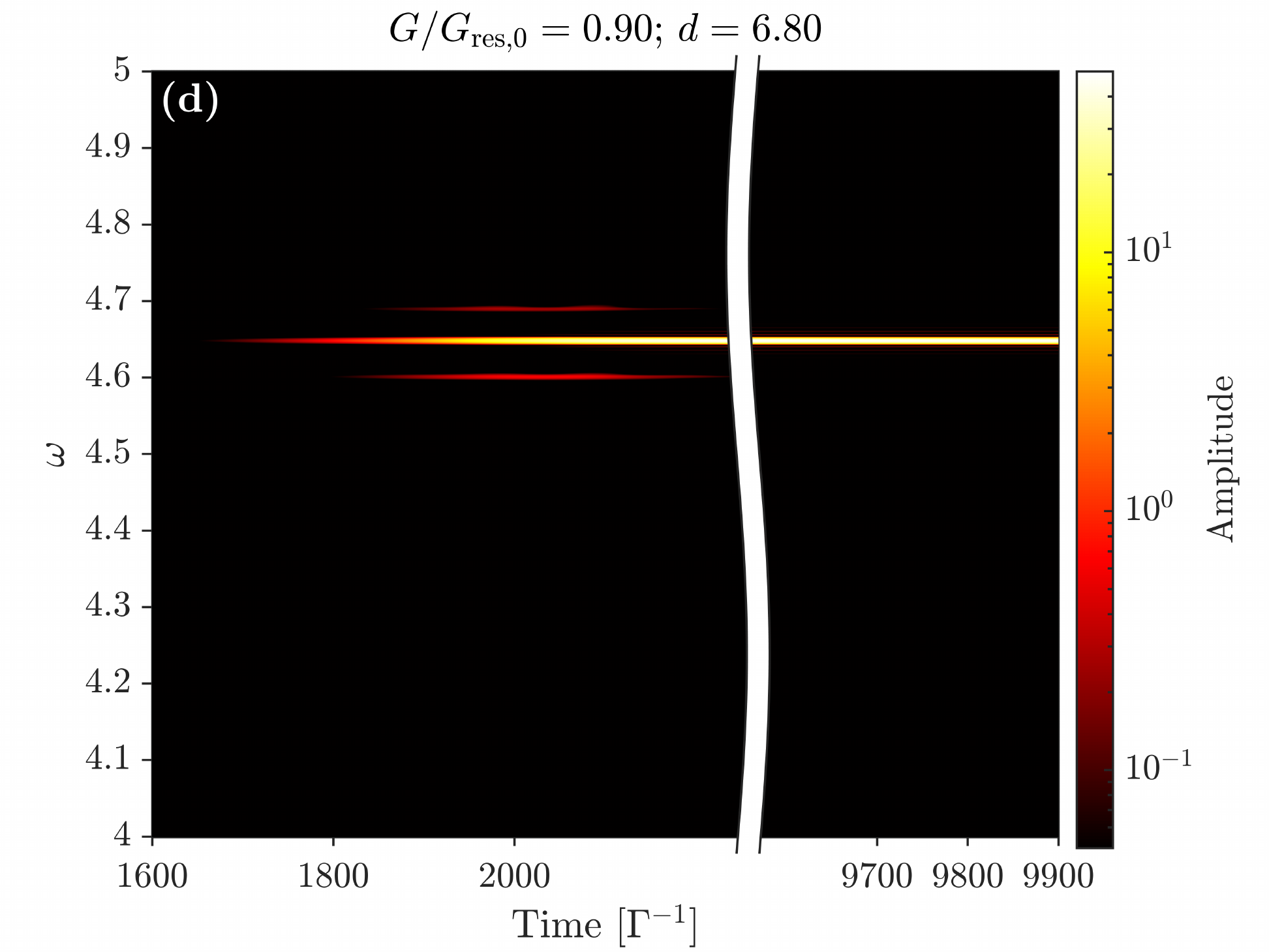}
        \hspace*{-6pt}\includegraphics[scale=0.5,valign=b]{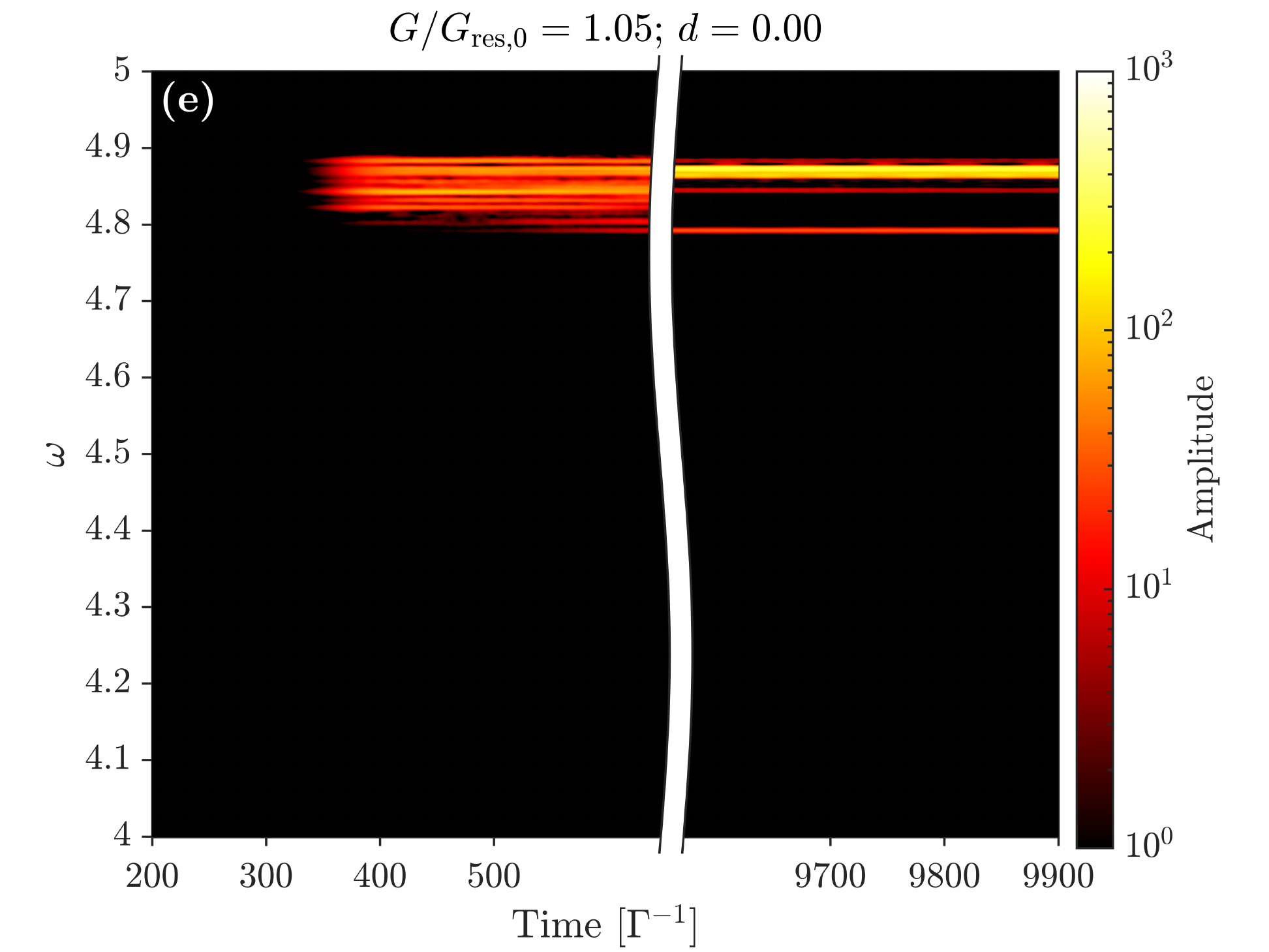}
    }
    \caption{Time-frequency representation (TFR) of the lasing process, for the same parameters used in Fig.~3(h),(g),(j) of the main text.
    Panels (a)--(b): vertically zoomed-in view of the band structure shown in Fig.~3(a)--(b) of the main text.
    Panels (c)--(e): TFR for the same parameters as in Fig.~3(h),(g),(j), respectively. The TFR is obtained by spatially averaging the smoothed pseudo Wigner-Ville distribution (WVD) of the field amplitude on individual lattice sites. The horizontal axis is broken so to better focus on the initial and final moments of the time evolution; the WVD close to the limits of the temporal intervals of the transform is also removed, as it displays well-known unphysical artifacts due to border effects. The amplitude of the WVD is shown in log scale, color-coded according to the colorbar on the right of each panel; the colorbars span three orders of magnitude, and the ones in panels (c) and (e) are matched for reading convenience.}
    \label{supp:fig:broadband_wvd}
\end{figure}

In this section we discuss the spectral features of the lasing emission for the scheme presented in Sec.~IV of the main text. As representatives for some of the different regimes identified in the phase diagram, we focus on three different parameter sets leading to the long-time snapshots of the emitted intensity shown in Fig.~3(h),(g),(j) of the main text. These parameters realize, respectively: a monochromatic topological laser, a multi-mode laser with a majority of the emission coming from a topological mode, and a non-topological multi-mode laser whose emission is due to the lasing of non-topological modes only.

The evolution of the system, numerically integrated up to $T=500\Gamma^{-1}$ for the three different cases, is shown in Fig.~\ref{supp:fig:broadband_wvd} as a time-frequency representation (TFR), where the color represents (in log scale) the spatially averaged amplitude of the smoothed pseudo Wigner-Ville distribution\cite{OToole2013} (WVD) of field amplitude on individual lattice sites.  In Fig.~\ref{supp:fig:broadband_wvd}(a)--(b) we focus around the relevant lasing frequencies for these cases; as a first observation, in the considered frequency interval the gain is effectively broadband (not shown). The band structure can also be used to classify the frequencies emerging from the TFR in panel (c), corresponding to a multi-mode laser in which, though, most of the emission is still coming from a topological edge mode. In this panel, the topological edge modes correspond to the well separated horizontal features of the WVD falling inside the frequency range of the topological bandgap shown in panel (a). There are striking signatures of two distinct effects: the frequency discretization discussed in Sec.~\ref{supp:sec:mode_pull}, and the well-known mode-competition. While due to the frequency discretization there are multiple topological modes that are able to ignite the lasing process, these modes spatially share the same active medium and therefore compete with each other until, at long enough times, only one of them is left.\cite{Sargent1974} The upper horizontal features, closer to each other, correspond instead to bulk modes located in the upper band of the central region; also in this case there are multiple modes that ignite the lasing process, but not all them survive at long times.

If we turn off the surrounding trivial region (i.e.\@ if we set $d=0$, panel (e)), the modes in the topological bandgap are not able to ignite the lasing process and lasing occurs from the top band modes only. Note that the switch-on happens at a longer time, due to the fact that a surrounding trivial region with $d=0$ creates an additional loss channel, especially for the modes that have a considerable overlap with the trivial region itself.

If instead we reduce $G/G_{\mathrm{res},0}$ below the lasing threshold of the bulk states (panel (d)) but we correspondingly increase $d$, we are able to prevent lasing from the bulk modes and only retain edge mode lasing at long times, thus obtaining a monochromatic topological laser. Due to the fact that the mechanism for providing the gain relies heavily, in this case, on the overlap of the edge mode tail with the trivial region (see Sec.~\ref{supp:sec:transition_lines}), the switch-on is much slower than the previous cases.

\section{Phase Diagrams for Different Values of the Gain Linewidth}
\label{supp:sec:extra_phase_diagrams}

\begin{figure}[htbp]
    \centering
    \mbox{
        \hspace*{-12pt}\includegraphics[scale=0.5,valign=b]{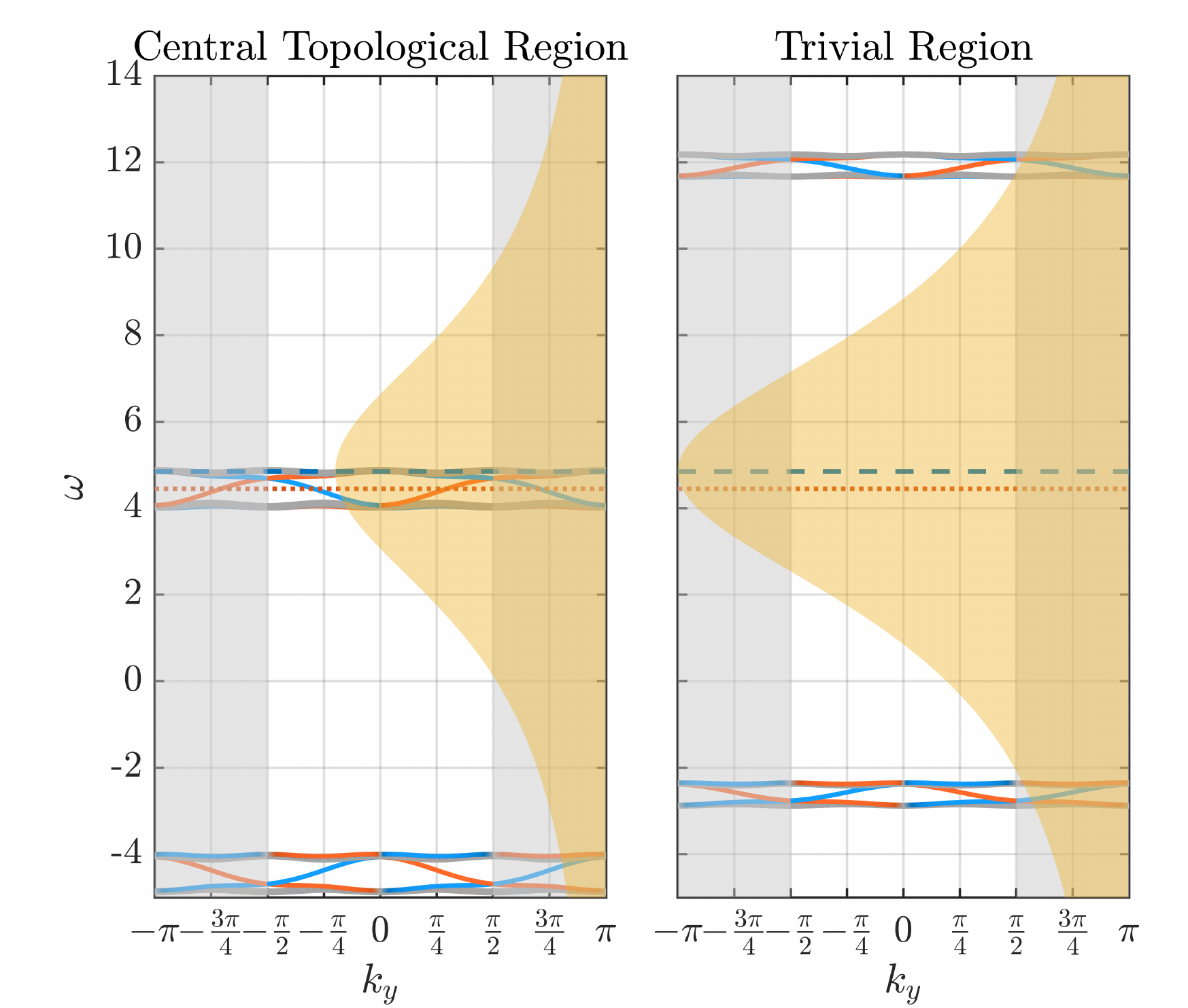}
        \hspace*{-12pt}\includegraphics[scale=0.5,valign=b]{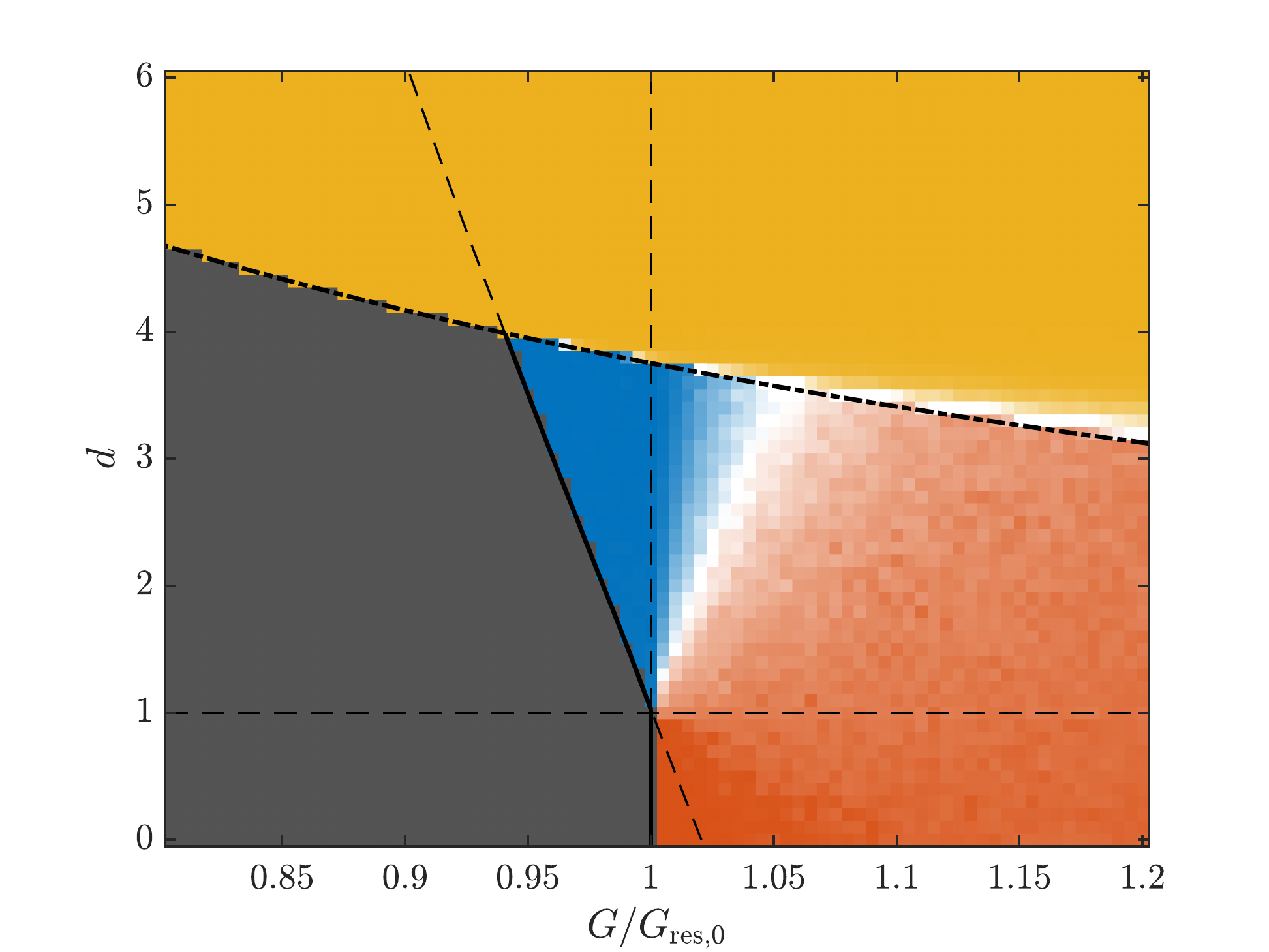}
    }
    \mbox{
        \hspace*{-12pt}\includegraphics[scale=0.5,valign=b]{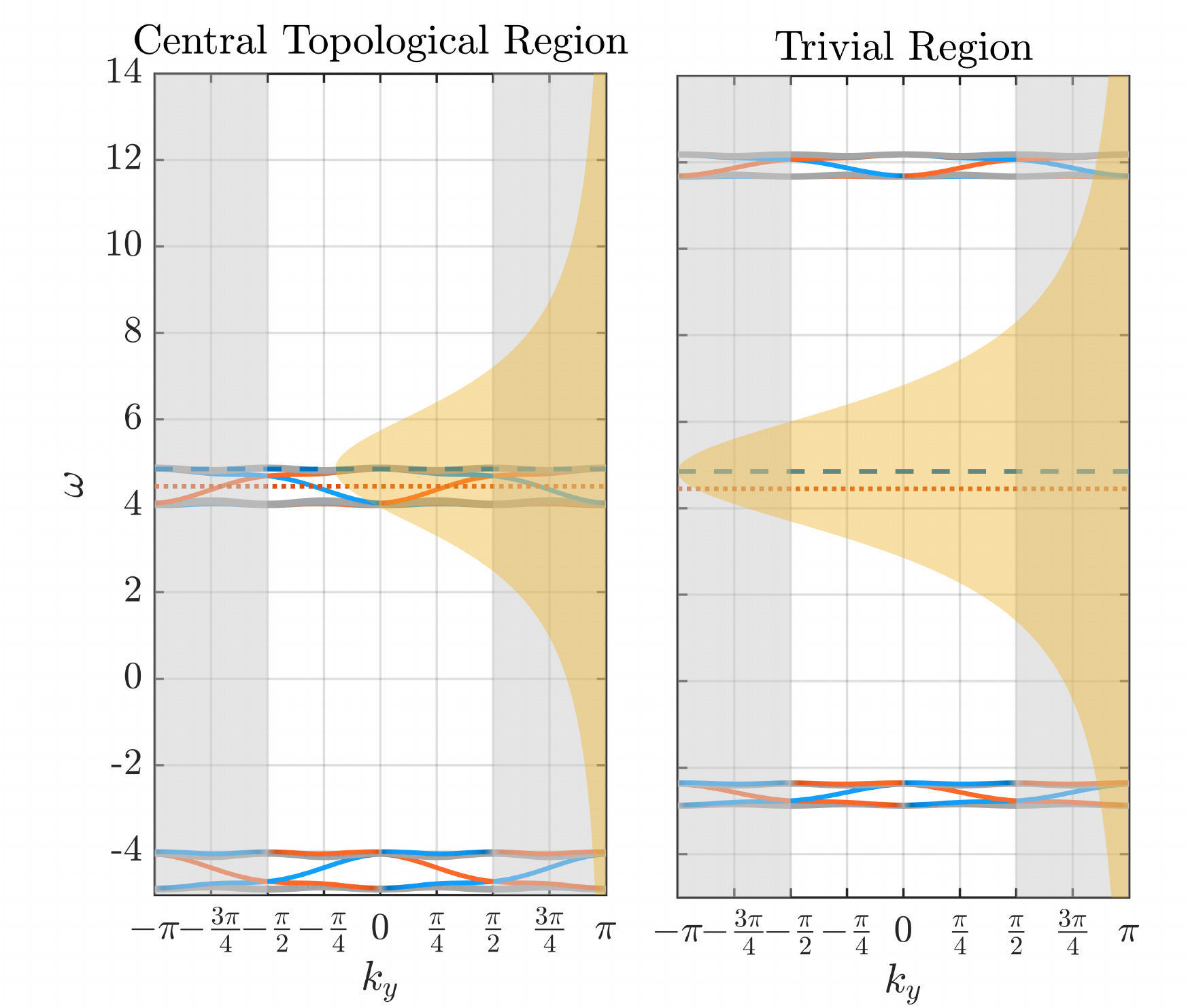}
        \hspace*{-12pt}\includegraphics[scale=0.5,valign=b]{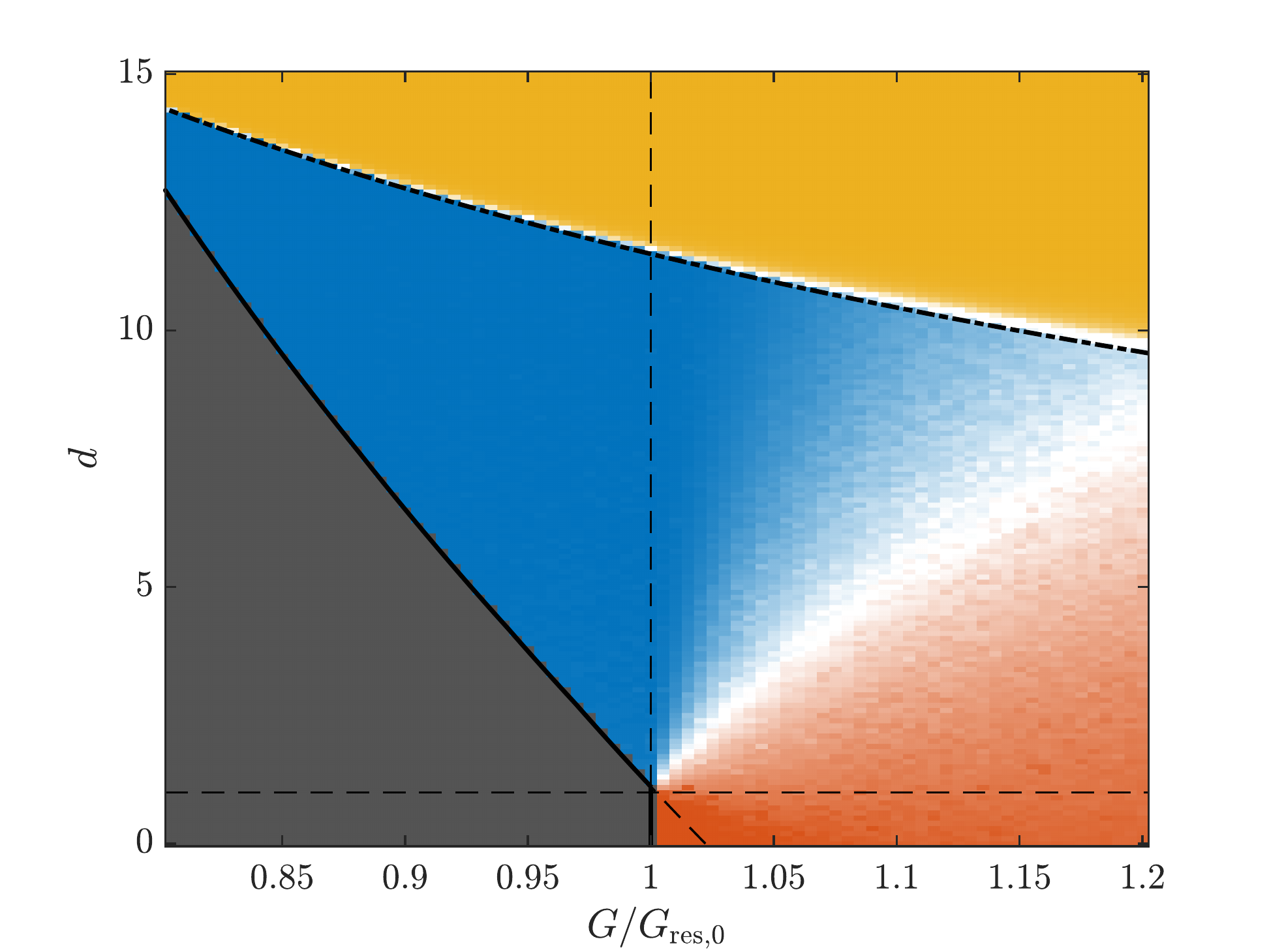}
    }
    \caption{Simulations obtained on a $25\times25$ lattice with a $5$-sites thick surrounding region with a trivial gap.
    (Left column) Band structure of the lattice. The central region has $\Delta = 4.0$, while the surrounding region has $\Delta_{\mathrm{trivial}} = 7.0$ and $\omega_{\mathrm{trivial}} = 4.65$ (red, dotted line). The gain is centered at $\omega_{eg} = 4.85$ (blue, dashed line).
    (Right column) Phase diagram of the different lasing regimes obtained from multiple simulations of equations \eqref{supp:eq:topolaser_TLA_equations} up to a long time $T=10^4/\Gamma$ as a function of the density $d$ of the surrounding region and the gain strength $G$. A grey color indicates no lasing; a blue color indicates topological lasing from the edge mode; a red color indicates lasing from the non-topological portion of the central region; a yellow color indicates lasing from the surrounding region. Fading to white indicates the coexistence of multiple phases. The panels on the two rows correspond to different values of the FWHM of the gain lineshape (yellow shading in the left panels), $2\gamma = 8.0$ (top), $2\gamma = 4.0$ (bottom). The solid and dot-dashed black lines in the phase diagrams indicate the analytical predictions in Sec.~\ref{supp:sec:transition_lines} for the transition between the different regions. The dashed black lines indicate $G/G_{\mathrm{res},0} = 1$ and $d = 1$.}
    \label{supp:fig:broadband_results_extra}
\end{figure}

We show here additional simulations of the bipartite Harper-Hofstadter laser with the central topological region and the surrounding trivial region discussed in Sec.~IV of the main text. In particular, we have pointed out in the main text that changing the gain linewidth reflects in a change of the area of the topological lasing region in the phase diagram. In Fig.~\ref{supp:fig:broadband_results_extra} we show simulations with an increased (top row) or reduced (bottom row) gain linewidth when compared to the value used in the main text.

For the increased (reduced) gain linewidth we take $2\gamma = 8.0$ ($2\gamma = 4.0$), to be compared with the value $2\gamma = 5.2$ used in the main text. With these settings the gain linewidth is around $\times 12.5$ ($\times 6.3$) wider than the linewidth of the topological bandgap of the central region, so it is roughly $96$ ($48$) times wider with respect to the topological bandgap than the gain linewidth shown in the narrowband case. The gain linewidth is also around $57\%$ ($29\%$) the linewidth of the trivial bandgap in the surrounding region.

When using this increased/reduced gain linewidth, the area of the region of parameters that yields a topological laser shrinks/widens; in particular, the upper transition line (which describes the lasing threshold of the surrounding region) is extremely sensitive to the linewidth of the gain. It is possible to predict in advance the effects of such a reduced/increased gain linewidth on the phase diagram by a direct calculation of the transition lines --- see Sec.~\ref{supp:sec:transition_lines}.

\begin{figure}[htbp]
    \centering
    \mbox{
        \hspace*{-25.5pt}\includegraphics[scale=0.5,valign=b]{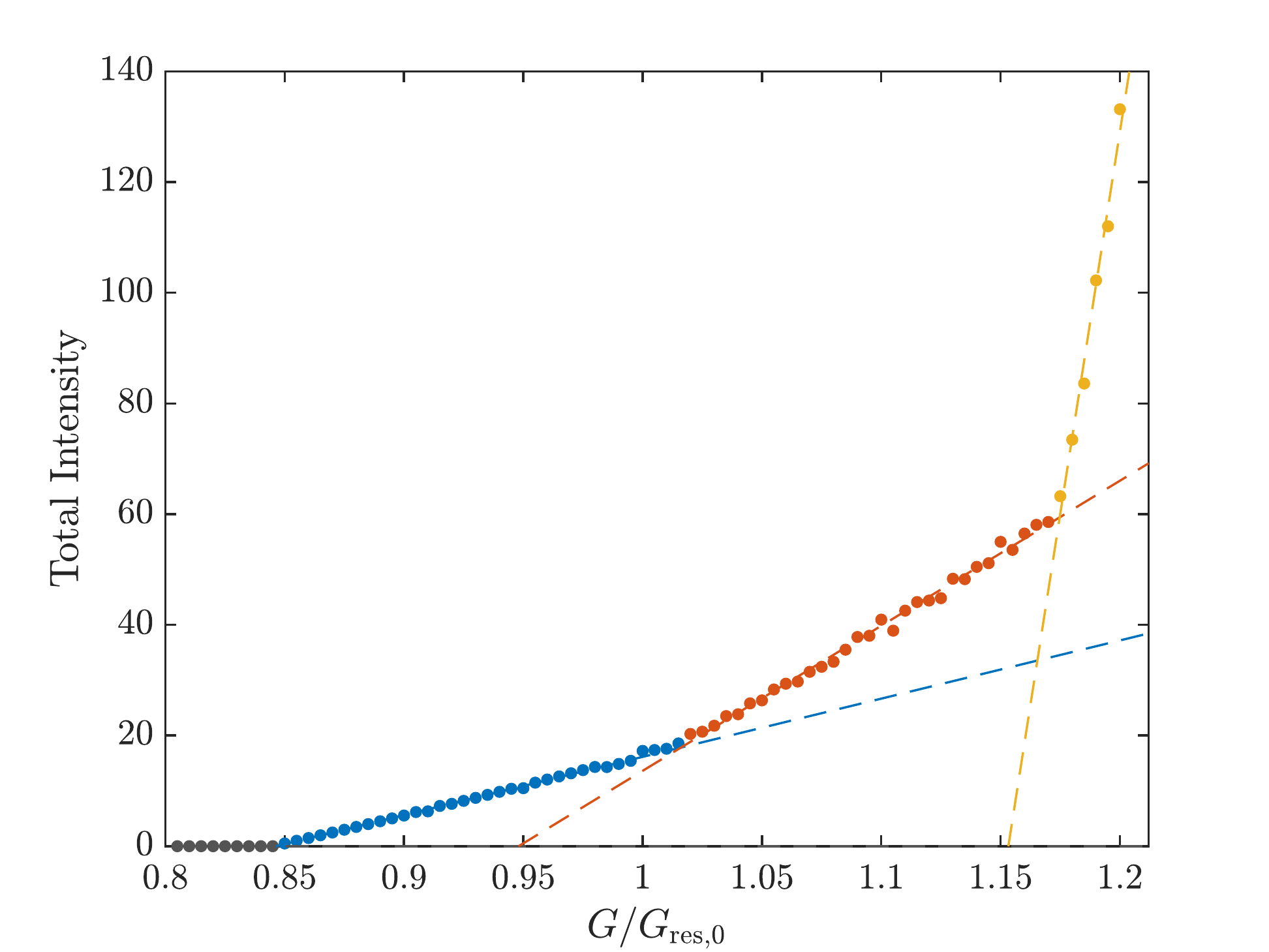}
        \hspace*{-22.5pt}\includegraphics[scale=0.5,valign=b]{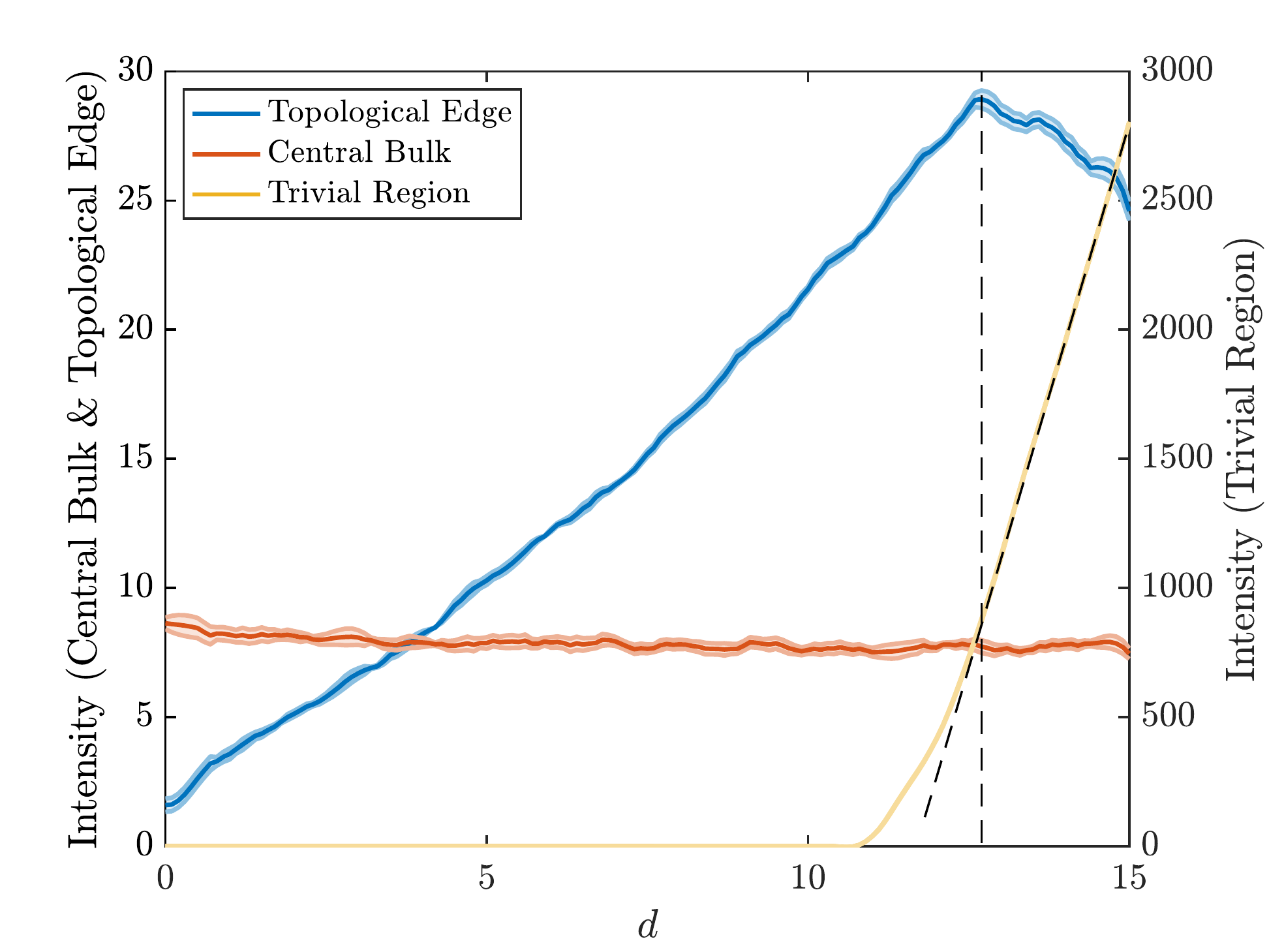}
    }
    \caption{Emitted intensity along a horizontal (left panel) and a vertical (right panel) cut of the phase diagram shown in the bottom right panel of Fig.~\ref{supp:fig:broadband_results_extra}.
    (Left panel) Total emitted intensity as a function of the gain strength $G/G_{\mathrm{res},0}$, at a fixed value of the density $d = 9.80$ of the surrounding region. The data has been classified in four color-coded groups, each fitted by a (accordingly colored) straight dashed line. Grey indicates no lasing, blue indicates lasing from the topological edge, red from the topological edge and the central bulk, and yellow from topological edge, central bulk and the surrounding region.
    (Right panel) Emitted intensity by each spatial region as a function of the density $d$ of the surrounding region, for a fixed value $G/G_{\mathrm{res},0} = 1.06$ of the gain strength. The intensity has been averaged in time for the last $1\%$ of the simulation, with the shaded areas corresponding to one standard deviation from the mean. The vertical dashed line marks the peak of the topological edge emission, while the oblique straight dashed line is a fit of the right-most part of the emission from the surrounding region.}
    \label{supp:fig:broadband_cuts_01}
\end{figure}

It is also instructive to explore the phase diagram, with respect to the emitted intensity, along both a vertical and a horizontal cut; we do that in Fig.~\ref{supp:fig:broadband_cuts_01} for the bottom right phase diagram of Fig.~\ref{supp:fig:broadband_results_extra}.

If we fix the effective density of gain material $d$ of the surrounding region at some value $d=9.80$ and we plot the emitted intensity vs $G/G_{\mathrm{res},0}$ (left panel of Fig.~\ref{supp:fig:broadband_cuts_01}) we notice that the data can be grouped in four sets, each one fitting a distinct linear branch. The fitted linear branches highlight the presence of three thresholds: the first one for the topological edge, the second one for the central bulk and the third one for the surrounding region.

We can also explore the emitted intensity along a vertical cut of the phase diagram, i.e.\@ as a function of the density $d$ of the surrounding region for a fixed value of the gain strength $G/G_{\mathrm{res},0} = 1.06$; for the sake of clarity, it is best to separate the emission from the topological edge region (blue), from the central bulk (red) and from the surrounding region (yellow). First of all we have to make a crucial observation: the scale of the emission from the surrounding region (right vertical axis) is two orders of magnitude bigger than the scale of the topological edge and of the central bulk (left vertical axis). This means that if we look at real-space snapshots of the system when the surrounding region is lasing (e.g.\@ Fig.~3(e)--(f) in the main text) we get the wrong impression that the rest of the system (i.e.\@ the central bulk and topological edge) is not lasing; the rest of the system \emph{is} indeed lasing, but the value of the emitted intensity is negligibly small when compared to the one emitted from the surrounding region. 

The other crucial observation is that the plot shows a clear evidence that the topological edge is indeed lasing by exploiting the unused amplification in the surrounding region. First of all, after the surrounding region is above threshold its emitted intensity has a non-linear ramp-up as a function of the density. This is due to the fact that, as one increases the density of the surrounding region, more and more modes from the surrounding region reach the lasing threshold. The first ones to lase are localized in the bulk of the surrounding region; however, at higher densities, the modes from the surrounding region localized on the border reach the lasing threshold as well. These modes, due to their sizable shared overlap, are in direct competition with the topological edge mode that is lasing in the central region; thus the emitted intensity of the latter starts decreasing (marked by a vertical dashed line in the plot).

\section{Transition Lines in the Phase Diagram}
\label{supp:sec:transition_lines}

As pointed out in Sec.~IV of the main text, the topological edge mode is able to lase thanks to an amplification leak from the surrounding region. This leak can be quantified and used to analytically determine the transition lines of the phase diagrams in Fig.~\ref{supp:fig:broadband_results_extra} and in Fig.~3(c) of the main text without having to resort at all to dynamical simulations.

The first ingredient is the knowledge of the energy spectrum and of the spatial structure of the eigenmodes of the lattice Hamiltonian, which we call $H_{\mathrm{bare}}$:
\begin{equation}
    H_{\mathrm{bare}}\psi_i = E_i\psi_i,
    \label{supp:eq:lattice_eigensystem}
\end{equation}
where $E_i$ is the energy corresponding to the mode with normalized wavefunction $\psi_i$, and $\psi_i$ can be thought as a matrix in position space, i.e.\@ $\psi_i = \left(\psi_{i;m,n}\right)$, with $m$ and $n$ respectively the $x$- and the $y$-index. 

\begin{figure}
    \centering
    \hspace*{-18pt}\includegraphics[scale=0.5]{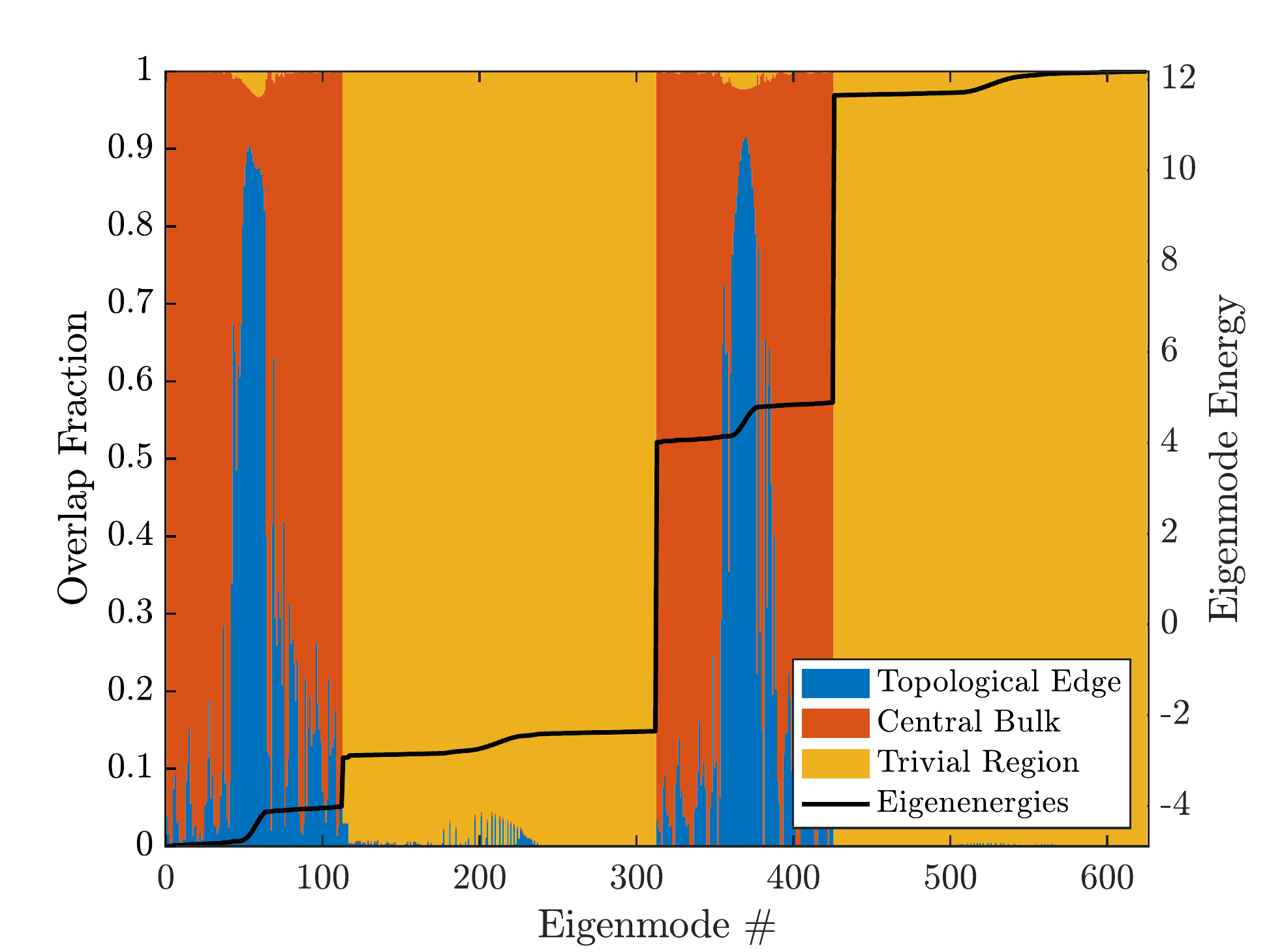}
    \caption{Overlaps $S_{\mathrm{topo}}$, $S_{\mathrm{bulk}}$ and $S_{\mathrm{trivial}}$ of each eigenmode with the different regions (colored shading, left vertical axis) and energies $E_i$ of the eigenmodes (solid black line). The calculations have been performed on the same $25\times25$ lattice geometry considered in Sec.~IV of the main text.}
    \label{supp:fig:overlaps_and_energies_03}
\end{figure}

At this point we can define some spatial masks $M_{\mathrm{topo}}$, $M_{\mathrm{bulk}}$ and $M_{\mathrm{trivial}}$; these masks are matrices in position space and contain $1$ in the entries representing points belonging to, respectively, the topological edge, the bulk of the central region and the surrounding region, and zero elsewhere. These masks are the same ones used to calculate the emitted intensity overlaps in the phase diagram, and also allow us to calculate the overlap of a given wavefunction with these three spatial region of the system, e.g.\@ to calculate the overlap $S_{\mathrm{topo},i}$ of the wavefunction $\psi_i$ with the topological edge:
\begin{equation}
    S_{\mathrm{topo},i} = \sum_{m,n}\lvert\psi_{i;m,n}\rvert^2 \cdot M_{\mathrm{topo};m,n}.
    \label{supp:eq:overlap_topo}
\end{equation}
For a given wavefunction $\psi_i$ the overlaps with the three regions sum to $1$ since the wavefunction is normalized, i.e.\@ $S_{\mathrm{topo},i}+S_{\mathrm{bulk},i}+S_{\mathrm{trivial},i}=1$. 
The overlaps of each mode can be visualized as in Fig.~\ref{supp:fig:overlaps_and_energies_03}.

The overlaps are useful to determine what is the \emph{effective} amplification $G_{\mathrm{eff},i}$ experienced by a given mode $\psi_i$ as compared to the global parameter $G$ characterizing the gain strength. For a completely flat gain profile, if all the system regions have the same density of the bulk, then $G_{\mathrm{eff},i}/G = 1$ because the overlaps of any given wavefunction with the three regions of the system sum to $1$. However, if we change the density $d$ of the surrounding region, this ratio will be in general $G_{\mathrm{eff},i}(d)/G = f_i(d) = S_{\mathrm{topo},i}+S_{\mathrm{bulk},i}+d \cdot S_{\mathrm{trivial},i}$. If $d>1$, then $f_i(d)>1$ and it acts as an \emph{enhancement factor} for the amplification of that mode.

If the gain is not flat but has e.g.\@ a Lorentzian profile, as in our case, the expression of $f_i(d)$ in terms of wavefunction overlaps has to be modified. Namely, we have to add an extra Lorentzian factor
\begin{equation}
     \mathcal{L}_i = \frac{1}{1+\left(\frac{\tilde{E}_i-\omega_{eg}}{\gamma}\right)^2}
     \label{supp:eq:lorentzian_factor}
\end{equation}
that is equal to $1$ at zero energy detuning and falls to $0$ at large energy differences. Analogously to Sec.~\ref{supp:sec:mode_pull}, the energies $\tilde{E}_i$ are here corrected with respect to $E_i$ in order to take into account the mode pulling, i.e.\@
\begin{equation}
    \tilde{E}_i = \frac{E_i+\mathcal{S}\omega_{eg}}{1+\mathcal{S}},
    \qquad
    \mathcal{S} = \frac{\Gamma}{\gamma}.
    \label{eq_energies_mode_pulling}
\end{equation}
Here, the mode pulling correction has the effect of increasing the value of $\mathcal{L}_i$ as compared to a Lorentzian factor calculated with the non-pulled energies $E_i$. So, the overall enhancement factor will be
\begin{equation}
     f_i(d) = \frac{S_{\mathrm{topo},i}+S_{\mathrm{bulk},i}+d \cdot S_{\mathrm{trivial},i}}{1+\left(\frac{\tilde{E}_i-\omega_{eg}}{\gamma}\right)^2}.
     \label{supp:eq:enhancement_factor}
\end{equation}

With these ingredients, we are ready to calculate the transition line between the not-lasing region and the different lasing regions, starting with the one corresponding to a topological laser. The first mode to lase, i.e.\@ the mode that determines the threshold, will be the mode with largest effective amplification, i.e.\@ with greater $f_i(d)$. That mode will have an enhancement factor equal to
\begin{equation}
    \bar{f}_{\mathrm{topo}}(d) = \max_{i\,:\,S_{\mathrm{topo},i}>\varepsilon_{\mathrm{topo}}} f_i(d).
    \label{supp:eq:topo_enhancement_final}
\end{equation}
Since we are looking for the lasing threshold among the topological edge modes, we take the $\max$ over the modes that have an overlap $S_{\mathrm{topo},i}$ with the topological edge greater than a certain threshold $\varepsilon_{\mathrm{topo}}$, e.g.\@ $\varepsilon_{\mathrm{topo}} = 0.80$. Similarly, to find the second transition line at which the surrounding region starts to lase, one defines
\begin{equation}
    \bar{f}_{\mathrm{trivial}}(d) = \max_{i\,:\,S_{\mathrm{trivial},i}>\varepsilon_{\mathrm{trivial}}} f_i(d).
    \label{supp:eq:pad_enhancement_final}
\end{equation}
Note that the mode that maximizes the enhancement factor depends on the specific $d$ being considered. A more crude approximation not involving a maximization over the modes, but that still provides very good results, would be to define e.g.\@ $\bar{f}_{\mathrm{topo}}(d) = f_{i_{\mathrm{topo}}}(d)$, where $i_{\mathrm{topo}}$ is the index of the topological edge mode that lases more frequently just above threshold in the time-domain simulations; $i_{\mathrm{topo}}$ is then independent on the choice of $d$. A similar consideration holds for $\bar{f}_{\mathrm{trivial}}(d)$ as well.

In general, the threshold for lasing into a given mode is reduced by the same factor $f_i$ characterizing the enhanced effective amplification as compared to the $d=1$ case for which the threshold is at $G/G_{\mathrm{res},0} = 1$. As a result, the transition lines we are looking for are simply described by the equations
\begin{equation}
    d = \frac{1}{\bar{f}_{\mathrm{topo}}(d)}
    \qquad\text{and}\qquad
    d = \frac{1}{\bar{f}_{\mathrm{trivial}}(d)}\,.
    \label{supp:eq:TL_equation_topo_pad}
\end{equation}
Both transition lines carry a dependency on $\gamma$, mediated by the Lorentzian factor in $f_i(d)$. However, the transition line between the not-lasing and the topo-lasing regimes has a much weaker dependence. The reason is that, by construction, the topological edge modes are much closer to the center of the Lorentzian than the bands in the surrounding region. Since this detuning contributes quadratically to the position of any transition line, the transition line involving modes farther from the gain center is therefore much more sensitive to $\gamma$ itself.

\begin{figure}
    \centering
    \mbox{
        \hspace*{-25.5pt}\includegraphics[scale=0.5,valign=b]{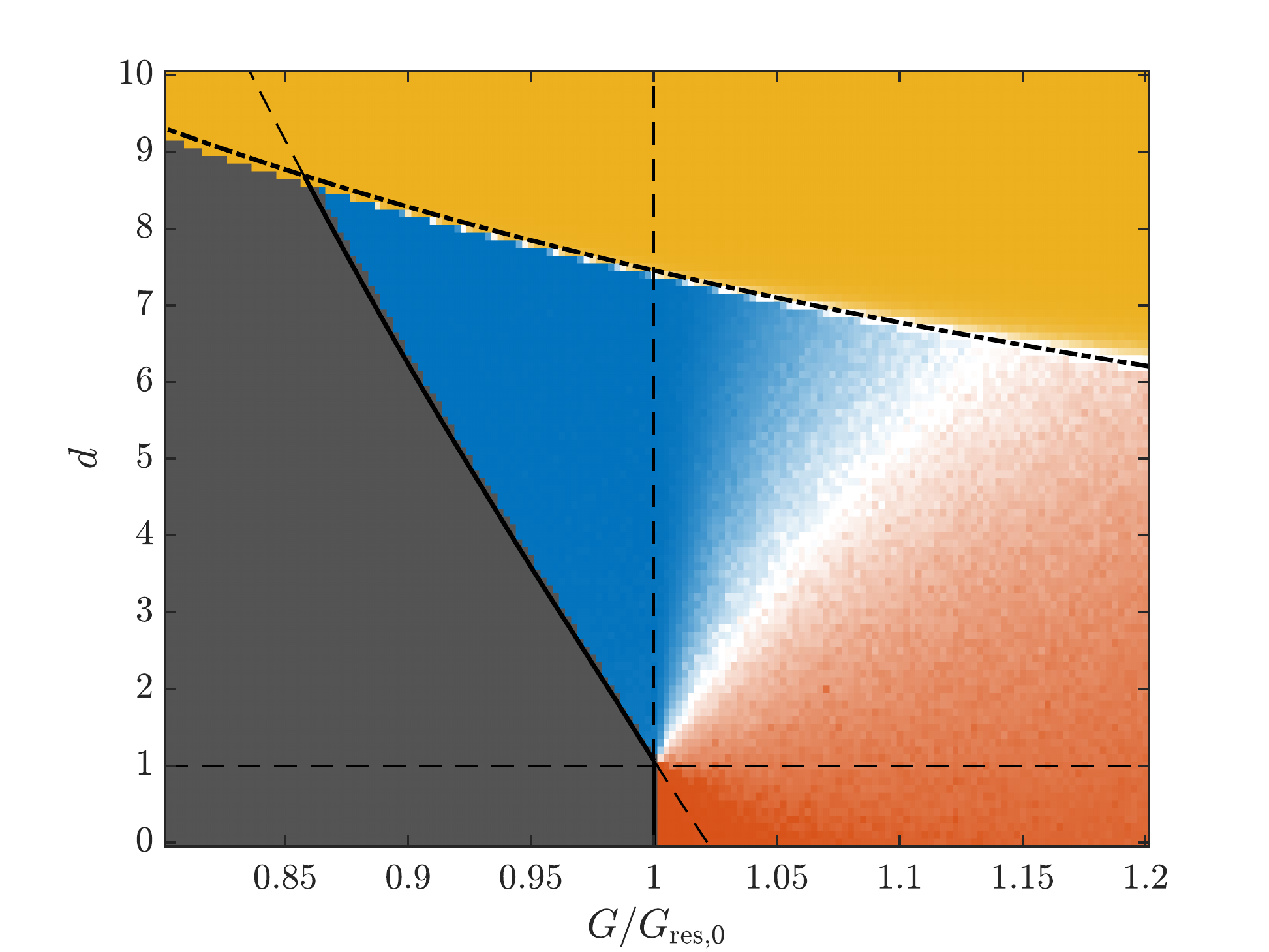}
        \hspace*{-22.5pt}\includegraphics[scale=0.5,valign=b]{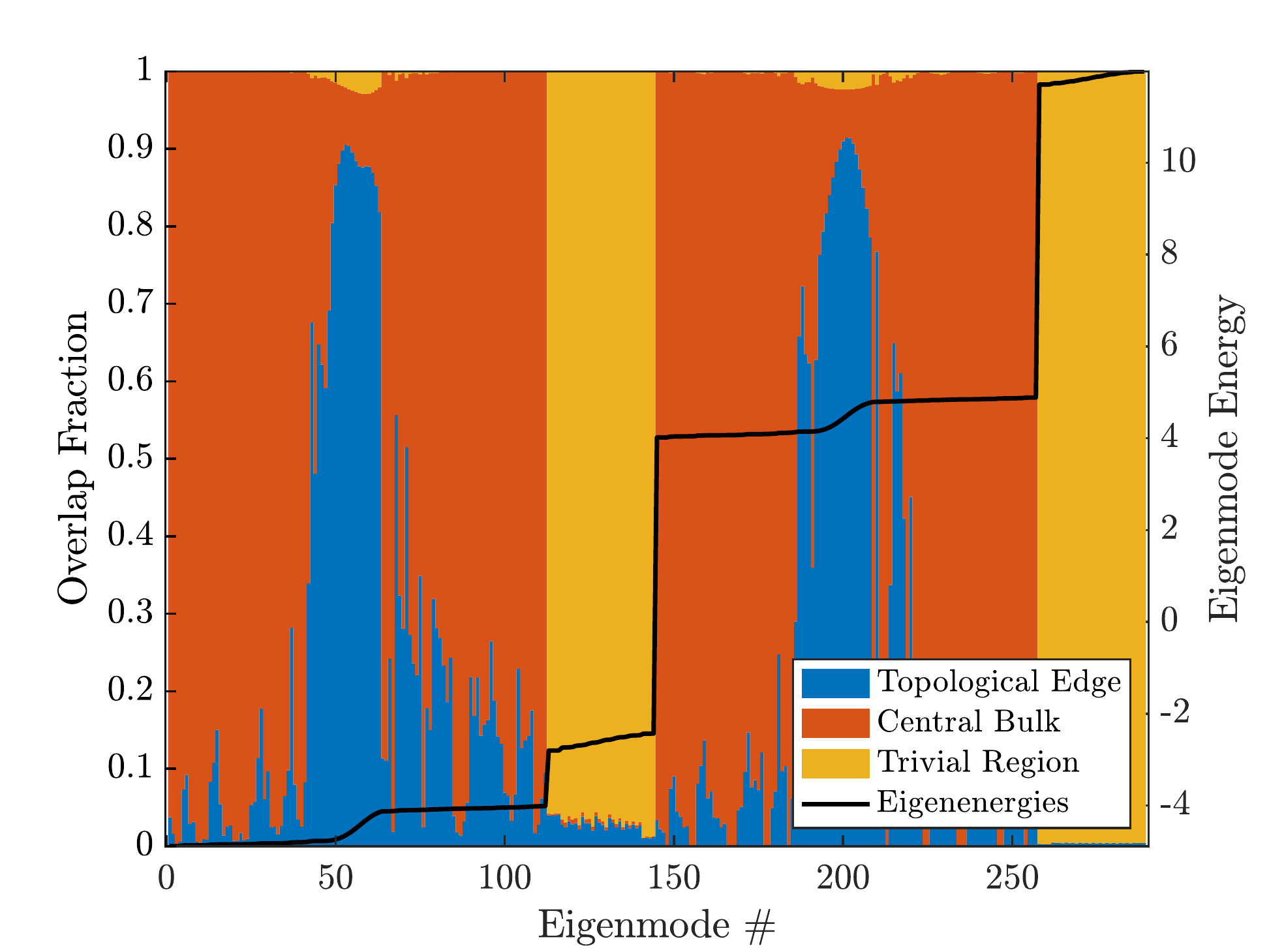}
    }
    \caption{Calculations performed on a $17\times17$ lattice geometry obtained from a $15\times15$ lattice surrounded by a 1-site-thick trivial region. All the other parameters are equal to the ones used in Fig.~3 of the main text.
    (Left) Transition lines of the considered $17\times17$ geometry with a 1-site-thick surrounding region on top of the phase diagram obtained with a 5-sites-thick surrounding region --- identical to the one shown in Fig.~3(c) of the main text.
    (Right) Overlaps $S_{\mathrm{topo}}$, $S_{\mathrm{bulk}}$ and $S_{\mathrm{trivial}}$ of each eigenmode with the different regions (colored shading, left vertical axis) and energies $E_i$ of the eigenmodes (solid black line).}
    \label{supp:fig:overlaps_and_energies_03_pad01}
\end{figure}

As a final note, this calculation also allows us to easily determine the lasing thresholds when we vary the thickness of the surrounding region. The system considered in Fig.~3 of the main text has a 5-sites-thick surrounding region; in the left panel of Fig.~\ref{supp:fig:overlaps_and_energies_03_pad01} we show again its phase diagram but we calculate the transition lines of an otherwise identical system which differs by having a 1-site-thick surrounding region. There is a discrepancy between the two, albeit extremely small, which shows that a bare minimum 1-site-thick surrounding region is already enough to open a topological lasing region in the phase diagram. Already with a 2-sites-thick surrounding region even this small discrepancy disappears, and the transition lines becomes virtually indistinguishable from the 5-sites-thick case.

The fact that a 1-site-thick surrounding region is already sufficient comes from the fact that the topological edge modes we want to enhance have an overlap with the surrounding region that is mainly concentrated in the surrounding sites closest to the edge itself, i.e.\@ in the system region covered by a 1-site-thick surrounding region. As can be inferred by comparing the overlap fractions shown in Fig.~\ref{supp:fig:overlaps_and_energies_03} and in the right panel of Fig.~\ref{supp:fig:overlaps_and_energies_03_pad01}, the positive-energy topological edge modes have an overlap with the surrounding region that's at best around $2.3\%$ both when the surrounding region is 1-site-thick and when it's 5-sites-thick, thus contributing around the same enhancing factor in the calculation of the thresholds.

\bibliographystyle{apsrev4-1}   
\typeout{}                      
\bibliography{references}